\documentclass[
    reprint,
    aps,
    pra,
    amsmath,
    amssymb,
    superscriptaddress,
    nofootinbib,
    longbibliography
]{revtex4-2}

\usepackage{mathtools}
\usepackage{physics}
\usepackage{graphicx}
\usepackage{bm}
\usepackage{xcolor}
\usepackage{dsfont}
\usepackage{array}
\usepackage{multirow}
\usepackage{enumitem}
\usepackage[edges]{forest}
\usepackage{tikz}
\usetikzlibrary{trees,positioning}
\usepackage{adjustbox}
\usepackage{qtree}
\usepackage{xspace}
\usepackage{hyphenat}
\usepackage{tabularx}
\usepackage{comment}
\usepackage{lipsum}

\definecolor{rossocorsa}{rgb}{0.83,0.0,0.0}
\definecolor{bleudefrance}{rgb}{0.19,0.55,0.91}

\usepackage{hyperref}
\hypersetup{
    colorlinks=true,
    filecolor=bleudefrance,
    linkcolor={green!50!blue},
    citecolor={red!50!black},
    urlcolor={blue!50!black}
}

\numberwithin{equation}{section}

\newcommand{\mb}{\mathbb}
\newcommand{\mc}{\mathcal}
\newcommand{\mr}{\mathrm}
\newcommand{\on}{\operatorname}
\newcommand{\p}{\partial}
\newcommand{\f}{\frac}

\allowdisplaybreaks[4]

\begin{document}
\title{Conjugate measurement-induced evolution towards maximally mixed states in quadratic quantum systems}

\author{S. Adarsh}
\affiliation{Department of Physics, National Institute of Technology Calicut, Kozhikode 673601, India.}

\author{T. P. Sreeraj}
\affiliation{Department of Physics, National Institute of Technology Calicut, Kozhikode 673601, India.}

\begin{abstract}
    In an earlier work, we showed that, in the interaction-dominant setting, coupling the conjugate variables of a subsystem to distinct environment modes drives the subsystem to a maximally mixed state. Here, we extend the study to the full quadratic dynamics of a single-mode subsystem coupled to a two-mode environment, and identify the phase-space mechanism responsible for the asymptotic behavior. By tracking the exact reduced Weyl characteristic and Wigner functions, we demonstrate that they lose their dependence on local phase-space coordinates on every fixed compact region. In free-particle model, this happens for all values of coupling, whereas in the harmonic oscillator model, it happens above a critical value of the coupling. Under a finite-volume and momentum regularization, this behavior corresponds to the reduced subsystem approaching a maximally mixed state.
\end{abstract}

\maketitle

\section{Introduction}
\label{introduction}

A fundamental challenge in quantum statistical mechanics and open quantum systems theory is explaining how equilibrium-like behavior arises from strictly unitary dynamics. The standard open-system setting addresses this by coupling a distinguished subsystem to an environment. The environment degrees of freedom are integrated out to obtain a reduced description of the system \cite{FeynmanVernon1963}. In appropriate limits, this reduction leads to Markovian evolution governed by Gorini--Kossakowski--Sudarshan--Lindblad master equations \cite{Gorini1976, Lindblad1976}. An alternative question is how the equilibration can emerge within the full unitary framework. Within this framework, mechanisms such as canonical typicality \cite{Popescu2006,Goldstein2006}, phase relaxation, and dynamical relaxation account for the emergence of stationary subsystem behavior \cite{Reimann2008, Linden2009, ShortFarrelly2012, GogolinEisert2016}.

In our earlier work \cite{Adarsh:2026btz}, we considered a minimal subsystem--environment model, in which the conjugate variables $x$ and $p$ of the subsystem are coupled to two independent environment degrees of freedom through a von Neumann-type interaction. We studied the interaction-dominated regime of this model and found that the coupling simultaneously suppresses coherence in both the conjugate bases of the subsystem, asymptotically leading to a maximally mixed description. This suggested a direct connection between conjugate measurements, asymptotic equilibration, and the emergence of equal a priori probabilities.

The earlier analysis, however, left several questions open. Its main argument was developed in the regime where the interaction dominates the intrinsic Hamiltonians of the system and the environment modes. The effect of the latter was addressed only through a sufficient condition. It therefore remained to be determined whether the behavior would survive when the free Hamiltonian of the component modes are retained and what dynamical structure produces the asymptotic behavior.

Standard formulations of quantum equilibration typically rely on large effective Hilbert-space dimensions, extended many-body degrees of freedom, or infinite thermal reservoirs. However, nontrivial equilibration of a reduced subsystem can also occur in coupling to finite quantum environments \cite{GemmerMichel2006, GemmerMichel2006EPJB}. Our earlier work \cite{Adarsh:2026btz} provided a particularly minimal setting with two environmental modes. The question addressed in this work is, therefore, not whether a finite number of modes can produce equilibration-like behavior, but whether such behavior persists under the exact quadratic evolution and what phase-space mechanism is responsible for it.

Quadratic Hamiltonians offer an ideal setting to investigate this problem. Under quadratic dynamics, the Heisenberg evolution of canonical phase-space variables remains strictly linear and is described by a real symplectic transformation \cite{SimonMukundaDutta1994, ArvindDuttaMukundaSimon1995, MukundaSimonSudarshan1988, Weedbrook2012, Serafini2017}. In the Wigner--Weyl formulation of quantum mechanics \cite{Wigner1932, Moyal1949, Hillery1984}, the evolution can be followed directly in phase space. This provides a clear geometric picture of the dynamics that remains valid for arbitrary, non-Gaussian initial states.

Local relaxation under quadratic Hamiltonians has been extensively studied in extended bosonic and fermionic lattice models. Such studies have demonstrated exact local relaxation, the Gaussification of non-Gaussian initial states, and relaxation toward Gaussian or generalized Gibbs ensembles \cite{CramerDawsonEisertOsborne2008, CramerEisert2010, GluzaKrumnowFriesdorfGogolinEisert2016, GluzaEisertFarrelly2019, MurthySrednicki2019, MonnaiMorodomeYuasa2019}. These phenomena generally rely on delocalizing transport, clustering properties, or thermodynamic-limit structures. In the present work, we return to the three-mode problem of our earlier study, but retain the full quadratic dynamics of the subsystem and the environment modes.

This framework uncovers two distinct physical mechanisms for the reduced asymptotic behavior. In the free-particle model, the dynamics produces a polynomial phase-space shear. Any nonzero argument of the Weyl characteristic function describing the system is transported to arbitrarily large phase-space distances. Since the Weyl characteristic function vanishes at infinity, the corresponding Weyl components are suppressed.

The harmonic oscillator model provides the main extension beyond the interaction-dominated analysis. For an analytically tractable choice of masses, frequencies, and coupling strengths, the dynamics undergoes a qualitative change when
\begin{equation}
	g_c^2 = \frac{32}{27}\omega^2.
\end{equation}
The spectrum of the generator transitions from purely oscillatory motion to a hyperbolic regime featuring exponentially growing and decaying modes. Such hyperbolic instabilities are already known in quadratic systems and symplectic geometry \cite{Kustura2019, Subramanyan2021}. Provided the initial Wigner function satisfies basic regularity conditions, it becomes locally flat at long times. Thus, in the harmonic oscillator case, unlike the phase-space shear seen in free-particle dynamics, the hyperbolic nature of the Hamiltonian spectrum is responsible for the Wigner flattening.

Together, these two models show that the phase-space dependence of the reduced Wigner function can disappear through different mechanisms: polynomial shear in the free model and hyperbolic expansion and contraction in the oscillator model. A globally uniform Wigner function over a non-compact phase space cannot be normalized. We resolve this by extending the finite-volume and a momentum cutoff that was already established in \cite{Adarsh:2026btz}. Within this regularized, finite-dimensional Weyl–Wigner framework \cite{Schwinger1960, Wootters1987, GalettiToledoPiza1988, Vourdas2004}, the asymptotic picture takes a clear physical meaning: the reduced subsystem asymptotically approaches the maximally mixed state.

Our goal here is not to derive standard thermalization from a small heat bath. We do not assume a background Gibbs ensemble, temperature, or detailed balance. Instead, we isolate a specific phase-space mechanism that allows a subsystem of a closed, finite-mode system to lose its local phase-space coordinate dependence. At the same time, the overall dynamics remains strictly unitary and reversible. Beyond offering an exactly solvable model for this kind of reduced equilibration, this analysis suggests a clear path forward: extending the framework to more general quadratic environments, such as a continuous reservoir of harmonic oscillators defined by a spectral density.

\section{General mechanism of Symplectic evolution}
\label{general_mechanism}

\subsection{Quadratic bipartite dynamics}

In the following sections, we consider the extension of the interaction used in \cite{Adarsh:2026btz}, by embedding the von Neumann-type interaction in the intrinsic Hamiltonians of the three modes involved. But, now, we first formulate the reduced dynamics of a general three-mode system independently of the particular model considered below. Let a closed quadratic system be partitioned into a distinguished subsystem $S$ and an environment $E$, with a canonical phase-space vector
\begin{equation}
    R= \begin{pmatrix}
        x & p &x_1 & x_2 & p_1 & p_2
    \end{pmatrix}^T,
\label{canonical_variables}
\end{equation}
where we define
\begin{equation}
R_S\equiv \begin{pmatrix}
        x\\
        p
    \end{pmatrix}, \qquad R_E \equiv \begin{pmatrix}
       x_1 & x_2 & p_1 & p_2
    \end{pmatrix}^T.
\end{equation}
A quadratic Hamiltonian can be written as
\begin{equation}
    H=\frac{1}{2}R^{T}GR,
\end{equation}
where $G=G^{T}$. The Heisenberg equations are linear,
\begin{equation}
\begin{gathered}
    \dot R(t)=MR(t),\\
    M=\Omega G, \qquad \Omega =J\oplus \begin{pmatrix}
    0 & \mathbf 1_2\\
    -\mathbf 1_2 & 0
\end{pmatrix},
    \end{gathered}
\end{equation}
where $J = \begin{pmatrix}
    0 & 1\\
    -1 & 0
\end{pmatrix}$ and $\mathbf 1_2$ represents the identity matrix of dimension $2\times 2$. The exact evolution is
\begin{equation}
    R(t)=S(t)R(0),
    \qquad
    S(t)=e^{Mt}.
\end{equation}
The propagator $S(t)$ is symplectic, in our case belonging to $Sp(6,\mb R)$ \cite{ArvindDuttaMukundaSimon1995, MukundaSimonSudarshan1988, Weedbrook2012,Serafini2017},
\begin{equation}
    S^{T}(t)\Omega S(t)=\Omega,
    \qquad
    \det S(t)=1.
\end{equation}
We partition $S(t)$ according to the subsystem--environment decomposition,
\begin{equation}
    S(t)=
    \begin{pmatrix}
        A(t) & B(t)\\
        C(t) & D(t)
    \end{pmatrix},
    \label{S_block_general}
\end{equation}
where $A(t)$ is the $2\times2$ block mapping initial subsystem coordinates into the final subsystem phase space.

\subsection{Exact reduced Wigner dynamics}\label{reduced_wigner_dynamics}

We describe the quantum state in phase space using Weyl displacement operators and their corresponding characteristic functions. For a canonical phase-space vector $R$, the Weyl operator associated with a phase-space displacement $\xi$ is given by
\begin{equation}
    \hat D(\xi)=\exp\left(\frac{i}{\hbar}\xi^{T}\Omega \hat R\right).
    \label{Weyl_displacement_definition}
\end{equation}
The hat on $\hat R =\begin{pmatrix}\hat x & \hat p &\hat x_1 & \hat x_2 & \hat p_1 & \hat p_2\end{pmatrix}^T$ is to make clear the fact that it is now an operator acting on the Hilbert space of quantum states. The Weyl characteristic function of a density operator $\rho$ is then
\begin{equation}
    \chi(\xi)=\on{Tr}\!\left[\rho \exp\left(\frac{i}{\hbar}\xi^{T}\Omega \hat R\right)\right].
    \label{Weyl_characteristic_definition}
\end{equation}
The characteristic function contains the same information as the density operator. The inverse of the above relation is
\begin{equation}
\rho =\frac{1}{(2\pi\hbar)^{n}}\int _{\mathbb{R}^{2n}}\,d^{2n}\xi\,\chi (\xi )\exp\left(-\frac{i}{\hbar}\xi^{T}\Omega \hat R\right),
\label{rho_in_terms_of_chi}
\end{equation}
where the integration is over the phase-space $\mathbb R^{2n}$, and $n=3$ corresponding to the three modes in our case. The reduced density matrix of the system can be obtained by setting the $\xi_E = 0$, and restricting the integral to only the system variables as
\begin{equation}
\rho_S =\frac{1}{(2\pi\hbar)}\int _{\mathbb{R}^2}\,d^{2}\xi_S\,\chi (\xi_S, \xi_E = 0 )\exp\left(-\frac{i}{\hbar}\xi_S^{T}\Omega_S \hat R_S\right),
\label{reduced_rho_in_terms_of_chi}
\end{equation}
The Weyl characteristic function is linked directly to the Wigner function through a symplectic Fourier transform \cite{Hillery1984, MukundaSimonSudarshan1988, Weedbrook2012}
\begin{equation} 
	W(R) = \frac{1}{(2\pi\hbar)^{2n}} \int_{\mathbb{R}^{2n}} d^{2n}\xi\, \chi(\xi)\, \exp\!\left( -\frac{i}{\hbar}\xi^{T}\Omega R \right).
\label{Wigner_from_characteristic} 
\end{equation}
Note that in the above expression $R$ is a vector of numerical values representing the phase-space coordinates.

For a quadratic Hamiltonian, the Heisenberg evolution of the canonical variables is linear,
\begin{equation}
    U^{\dagger}(t) \hat R U(t) = S(t)\hat R,
\end{equation}
where $S(t)$ is symplectic. Therefore,
\begin{equation}
    U^{\dagger}(t)\hat D(\xi)U(t)=\exp\left[\frac{i}{\hbar}\xi^{T}\Omega S(t)\hat R\right].
\end{equation}
Using the symplectic relation
\begin{equation}
    S^{T}\Omega S=\Omega,
\end{equation}
we have
\begin{equation}
    \xi^{T}\Omega S = \left(S^{-1}\xi\right)^{T}\Omega.
\end{equation}
Hence,
\begin{equation}
    U^{\dagger}(t)\hat D(\xi)U(t) = \hat D\!\left(S^{-1}(t)\xi\right).
    \label{Weyl_symplectic_evolution}
\end{equation}
The characteristic function at time $t$ is consequently
\begin{equation}
\begin{split}
    \chi_t(\xi)&=\on{Tr}\!\left[\rho(t)\hat D(\xi)\right]\\
    &=\on{Tr}\!\left[\rho(0)U^{\dagger}(t)\hat D(\xi)U(t)\right]\\
    &=
    \chi_0\!\left(S^{-1}(t)\xi\right).
    \label{Weyl_characteristic_evolution}
\end{split}
\end{equation}
Thus, the characteristic function evolves by composition with the inverse symplectic map.

Since the Wigner function is the symplectic Fourier transform of the Weyl characteristic function, the corresponding evolution in phase space follows
\begin{equation}
    W_t(R) = W_0\!\left(S^{-1}(t)R\right).
    \label{Wigner_symplectic_flow}
\end{equation}
These relations make no assumptions about the initial states of the system, and are valid for quadratic Hamiltonian dynamics.

Tracing over the environmental variables gives the reduced Wigner function
\begin{equation}
    W_S(R_S,t)=\int dR_E\,W_0\!\left[S^{-1}(t)
        \begin{pmatrix}
            R_S\\
            R_E
        \end{pmatrix}
    \right].
    \label{reduced_Wigner_start}
\end{equation}
At times for which $A^{-1}(t)$ is invertible, we may perform a linear change of environmental integration variables and obtain
\begin{equation}
    W_S(R_S,t)=\frac{1}{|\det A(t)|}\int d u\,W_0\left(A^{-1}(t)R_S-F(t)u,u\right),
    \label{reduced_Wigner_general}
\end{equation}
where
\begin{equation}
    F(t)\equiv A^{-1}(t)B(t).
    \label{F_definition}
\end{equation}
Details can be found in Appendix~\ref{reduced_wigner}. Equation~\eqref{reduced_Wigner_general} reveals the fact that the matrix $A^{-1}(t)$ controls how system coordinates $R_S$ enters the argument. Similarly, $F(t) = A^{-1}(t)B(t)$ determines how the environmental variables show up before being integrated out.

Below, we investigate two quadratic models that exhibit distinct asymptotic mechanisms. In the free-particle model, every nonzero phase-space point in the subsystem plane is asymptotically mapped by $S^{-1}(t)$ to points infinite distance way. Thus, the reduced Weyl characteristic function tends to zero at every nonzero argument. In the harmonic oscillator model, by contrast, $A^{-1}(t)$ tends to zero, making every nonzero separation in phase-space approach zero. Under the regularity assumption stated below, this gives local Wigner flattening.

\subsection{Definition of local Wigner flattening}

We now turn our attention to what exactly we mean by local Wigner flattening. A nonzero constant Wigner function is not normalizable over a noncompact continuum phase space. We therefore characterize the result as a change through the variation of the Wigner function within each compact regin of phase-space. For a compact region $K\subset\mathbb R^{2}$, define
\begin{equation}
    \Delta_K(t)=\sup_{R_S,R'_S\in K}\left|W_S(R_S,t)-W_S(R'_S,t)\right|.
    \label{DeltaK}
\end{equation}
We say that the reduced Wigner function undergoes local phase-space flattening when
\begin{equation}
    \Delta_K(t)\longrightarrow0
    \qquad
    \text{for every fixed compact }K.
    \label{flattening_definition}
\end{equation}
This gives an operational definition of the flattening mechanism. At long times, the difference between the reduced Wigner function evaluated at any two points within a fixed bounded region tends to zero. Equation~\eqref{flattening_definition}, however, should not be interpreted as convergence to a globally constant normalized Wigner function in the unregularized setting, as such a state does not exist on $L^{2}(\mathbb R)$. In Appendix~\ref{finite_volume_robustness}, we show that the flattening in the continuum setting reached over a fixed finite interval of time can be reproduced to controlled accuracy under sufficiently large spatial confinement. Convergence to a maximally mixed state can be shown under an additional momentum cutoff.

\section{Finite volume regularization}
\label{finite_volume_regularization}

We have encountered the idea of local Wigner flattening. The precise meaning of the phenomenon deserves more attention. A maximally mixed state, which is proportional to the identity $\mathbf{1}$, on an infinite-dimensional Hilbert space fails to be trace class. On $L^2(\mathbf R)$,
\begin{equation}\label{trace of identity}
    \on{Tr}\mathbf{1}=\infty.
\end{equation}
Hence, there exists no normalizable density matrix
\begin{equation}
    \rho_{\mathrm{mm}}\propto\mathbf{1}
\end{equation}
in the infinite-dimensional Hilbert space. In other words, a density operator proportional to the identity cannot be normalized on $L^2(\mathbf R)$. Thus, when we speak of the flattening of the Wigner function, it should not be understood as convergence to a normalizable constant. The regularization technique used here follows the construction of \cite{Adarsh:2026btz}. There the system is confined to a box of finite length, and a cutoff is imposed on the allowed momentum states. Here, we recast the construction to the finite-dimensional Wigner--Weyl representations.

The finite-dimensional construction introduced here is used only to trace the relationship between a discrete uniform Wigner function and a maximally mixed state. We do not introduce an independently truncated finite-dimensional Hamiltonian dynamics. Rather, after establishing the asymptotic behavior of the continuous-variable system, we represent its surviving Weyl components on the finite Hilbert space $\mc H_{L,N}$.

\subsection{Box regularization and momentum cutoff}

We introduce a finite-volume regularization. The system coordinate is restricted to
\begin{equation}\label{box regularization}
    x\in\left[-\frac{L}{2},\frac{L}{2}\right].
\end{equation}
Periodic boundary conditions, $\psi(x+L)=\psi(x)$, give the discrete momentum values,
\begin{equation}
    p_n=\frac{2\pi\hbar n}{L},
    \qquad
    n\in\mathbb{Z}.
    \label{box_momenta}
\end{equation}
We further introduce a momentum cutoff
\begin{equation}
    |n|\leq N.
    \label{momentum_cutoff}
\end{equation}
The regulated system Hilbert space is therefore
\begin{equation}
    \mc{H}_{L,N} = \on{span}\left\{\ket{p_n}:\,-N\leq n\leq N\right\},
\end{equation}
with finite dimension
\begin{equation}
    d_{L,N}=2N+1.
\end{equation}

In this setting, the maximally mixed state is well defined. In the momentum basis \cite{Vourdas2004},
\begin{equation}
    \rho_{\mathrm{mm}}^{(L,N)}=\frac{\mathbf{1}_{L,N}}{2N+1}=\frac{1}{2N+1}\sum_{|n|\leq N}\ket{p_n}\bra{p_n},
    \label{regulated_maximally_mixed}
\end{equation}
while in the position basis
\begin{equation}
    \rho_{\mathrm{mm}}^{(L,N)}=\frac{1}{2N+1}\int_{-L/2}^{L/2}dx\,d\bar{x}\,\delta_{L,N}(x-\bar{x})\ket{x}\bra{\bar{x}}.
    \label{mm_position_basis}
\end{equation}
Its trace is finite and normalized,
\begin{equation}
    \on{Tr}\rho_{\mathrm{mm}}^{(L,N)}=1.
\end{equation}
In Eq.~\eqref{mm_position_basis},
\begin{equation}
    \delta_{L,N}(x)=\frac{1}{L}\sum_{|n|\leq N}e^{i2\pi nx/L}=\frac{\sin\!\left[(2N+1)\pi x/L\right]}{L\sin(\pi x/L)}.
    \label{regulated_delta}
\end{equation}

The above construction converts the formal idea of a maximally mixed state in the infinite-dimensional theory into a well-defined, normalizable finite-rank operator. The same regularization also clarifies the meaning of Wigner flattening.

\subsection{Weyl modes and the uniform state} 
\label{discrete_Weyl_discussions}

Let $d=2N+1$ denote the dimension of the regularized Hilbert space introduced above. It is convenient to relabel the momentum states by $n\in\mb Z_d$, where $\mb Z_d$ represents the group of integers modulo $d$,
\begin{equation}
    \mb Z_d = \{0,1,2,\dots,d-1\}.
\end{equation}
The physical momentum associated with $n\in\mb Z_d$ is given by $2\pi\hbar \widetilde n/L$, where $\widetilde n \in[-N,N]$ and $\widetilde n \equiv n \,(\on{mod} d)$. On the finite Hilbert space $\mc H_{L,N}$, we define the unitary phase operator $Z$ and cyclic shift operator $X$ by
\begin{gather}
    Z\ket{p_n} = \omega_d^n\ket{p_n},
    \label{finite_Z}
    \\
    X\ket{p_n} = \ket{p_{n+1}},
    \label{finite_X}
\end{gather}
where $\omega_d = e^{2\pi i/d}$. Their continuum analogues are $Z(a)=e^{\frac{ia\hat p}{\hbar}}$, $X(b)=e^{\frac{ib\hat x}{\hbar}}$. The finite shift includes a cyclic identification at the momentum cutoff, thereby making it an analogue of the continuum momentum translation. In the finite Hilbert space under consideration they have the explicit forms
\begin{equation}
\begin{gathered}
Z=\sum_{n=0}^{d-1}\omega_d^n|p_n\rangle\langle p_n|,\\
X=\sum_{n=0}^{d-2}|p_{n+1}\rangle\langle p_n|+|p_0\rangle\langle p_{d-1}|.
\label{Z_and_X}
\end{gathered}
\end{equation}
The term $|p_0\rangle\langle p_{d-1}|$ makes the shift cyclic. They satisfy
\begin{equation}
    ZX = \omega_d XZ.
    \label{finite_weyl_relation}
\end{equation}

A complete set of finite Weyl operators is then
\begin{equation}
    D_{r,s}=\omega_d^{-\bar{2}rs} Z^r X^s,
    \qquad
    r,s\in\mathbb Z_d,
    \label{finite_weyl_operator}
\end{equation}
where $\bar{2}=\frac{d+1}{2}$ denotes the multiplicative inverse of $2$ modulo $d$, so that $2\bar{2}=1\pmod d$. Since $d=2N+1$ is odd, this inverse always exists. With the convention in Eq.~\eqref{finite_weyl_operator}, the displacement operators satisfy
\begin{equation}
    D_{r,s}^{\dagger}=D_{-r,-s},
\end{equation}
and
\begin{equation}
    D_{r,s}D_{r',s'} = \omega_d^{\bar{2}(rs'-sr')} D_{r+r',s+s'},
\end{equation}
where all indices are understood modulo $d$.

In particular,
\begin{equation}\label{zero_mode_Weyl_discrete}
    D_{0,0} = \mathbf 1_{L,N},
\end{equation}
The $D_{r,s}$ form an orthogonal basis of $\mc B(\mc H_{L,N})$, \cite{Schwinger1960,Vourdas2004},
\begin{equation}
    \on{Tr}\left(D_{r,s}^{\dagger}D_{r',s'}\right)=d\,\delta_{r,r'}\delta_{s,s'}.
    \label{finite_weyl_orthogonality}
\end{equation}
For a density operator $\rho_S^{(L,N)}(t)$ on the retained subspace,
\begin{equation}
    \chi_{r,s}(t)=\operatorname{Tr}\!\left[\rho_S^{(L,N)}(t)D_{r,s}\right].
\end{equation}
Using the orthogonality condition in Eq.~\ref{finite_weyl_orthogonality}, the density matrix can be expanded as 
\begin{equation}
    \rho_S^{(L,N)}(t)=\frac{1}{d}\sum_{r,s\in\mathbb Z_d}\chi_{r,s}(t)D_{r,s}^{\dagger},
    \label{density_weyl_expansion}
\end{equation}
where $\chi_{r,s}(t)$ is the discrete Weyl characteristic coefficients. Also, the discrete Wigner function is the finite symplectic Fourier transform of the Weyl coefficients \cite{GalettiToledoPiza1988, Vourdas2004},
\begin{equation}
    W_{a,b}^{(L,N)}(t)=\frac{1}{d^2}\sum_{r,s\in\mathbb Z_d}\chi_{r,s}(t)\omega_d^{\,as-br},
    \qquad
    a,b\in\mathbb Z_d.
    \label{discrete_wigner_fourier}
\end{equation}
See \cite{Wootters1987} for a foundational finite-state Wigner construction.

This provides the required formalism that helps establish the interpretation of the asymptotic results obtained below. In the continuum calculation below, the reduced Wigner function may progressively lose its dependence on the subsystem phase-space coordinates over any fixed compact region, while the distribution spreads over an increasingly large phase-space domain. In the regularized picture, the corresponding disappearance of the nonzero Fourier modes has an unambiguous interpretation: convergence to a maximally mixed state.

Before proceeding, we clarify that the regularization is introduced only to make the meaning of the asymptotic state precise and should not be regarded as a modification of the continuum dynamics. The regularization is used only after the continuum dynamics has been established. Thus, only after its asymptotic Weyl or Wigner structure has been determined do we represent that structure in the finite Weyl--Wigner space. In this representation, disappearance of all nonzero Weyl components leaves only the identity component, corresponding to the maximally mixed state. Only after this interpretation is made may one consider removing the regulators through $L,N\to\infty$.

\section{Free-particle model: shear-induced Weyl-mode suppression}
\label{free_particle_section}

In this section, we consider the free-particle model studied in \cite{Adarsh:2026btz}. There, the analysis was primarily done in the density matrix representation and identified the suppression of coherence in the conjugate bases. Here, we perform the same analysis in the symplectic picture and thereby show that the asymptotic behavior has a geometric origin in phase space. In this case, unlike the harmonic oscillator limit, discussed in the following section, $A^{-1}(t)$ does not asymptotically tend to a zero matrix. Instead, the interaction produces Weyl displacements whose environment coordinates increase polynomially with time. By using the box regularization introduced in the previous section, the asymptotic continuum state can be represented as the maximally mixed state of the regulated subsystem.

\subsection{Exact continuum dynamics}
\label{exact_continuum_dynamics_fp}

The Hamiltonian is of the form
\begin{equation}
    H_{\mathrm{fp}}=\frac{p^2}{2m}+\frac{p_1^2}{2m_1}+\frac{p_2^2}{2m_2}+g_1xp_1+g_2px_2,
    \label{free_hamiltonian}
\end{equation}
where $(x,p)$ denotes the subsystem and
$(x_1,p_1,x_2,p_2)$ the environmental degrees of freedom. 

Before proceeding with the dynamics, we note that the autonomous free-particle Hamiltonian is unbounded from below for any nonzero coupling. Rewriting $H_{\rm fp}$ by completing the squares gives
\begin{equation}
\begin{split}
H_{\rm fp}&=\frac{(p_1+m_1g_1x)^2}{2m_1}+\frac{(p+mg_2x_2)^2}{2m}+\frac{p_2^2}{2m_2}\\
&\quad
-\frac{m_1g_1^2}{2}x^2-\frac{mg_2^2}{2}x_2^2 .
\label{Hfp_completed_square}
\end{split}
\end{equation}
Thus, whenever $g_1 \neq 0$ or $g_2 \neq 0$, the quadratic form contains negative eigenvalues and remains unbounded from below. Consider, for example, choosing $p_1=-m_1g_1x$ and all other variables equal to zero gives $H_{\rm fp}=-(m_1g_1^2/2)x^2\to-\infty$ as $|x|\to\infty$. As a result, the free-particle model serves as an idealized effective quadratic interaction rather than as the globally stable Hamiltonian of three autonomous free modes. This energetic indefiniteness leaves the exact unitary symplectic evolution intact, but restricts the physical interpretation of the strict long-time limit. Two constructions help clarify the physical scope of this quadratic model. Firstly, analogous to the harmonic model, this quadratic interaction can be thought of as arising from a globally bounded nonlinear theory. Consider
\begin{equation}
H_{\rm fp}^{\rm full}=H_{\rm fp}+\frac{\lambda_1}{4}x^4+\frac{\lambda_2}{4}x_2^4,
\qquad
\lambda_1,\lambda_2>0.
\end{equation}
This is a Hamiltonian that is bounded from below, and whose expansion around the origin yields precisely the $H_{\rm fp}$. The free-particle model may therefore be interpreted as the local linear description in the neighborhood of an unstable stationary point of a nonlinear parent Hamiltonian. 

Secondly, in Appendix~\ref{finite_volume_robustness}, we provide a finite-volume construction where the spatial coordinates are confined within a box of finite size. This avoids the problem of unboundedness from below, yet reproduces the continuum dynamics with prescribed accuracy for any finite time interval and fixed box size. The argument is built around the harmonic oscillator model, but could be applied to the free-particle case as well. 

Returning to the exact, unconfined quadratic dynamics, the equations of motion are
\begin{align}
    \dot{x}&=\frac{p}{m}+g_2x_2,&\dot{x}_1&=\frac{p_1}{m_1}+g_1x,&\dot{x}_2&=\frac{p_2}{m_2},
    \label{free_x_eom}
    \\
    \dot{p}&=-g_1p_1,&\dot{p}_1&=0,&\dot{p}_2&=-g_2p.
    \label{free_p_eom}
\end{align}
The momentum equations can be integrated directly to give
\begin{gather}
    p_1(t)=p_1(0),
    \label{p1_free_solution}
    \\
    p(t)=p(0)-g_1t\,p_1(0),
    \label{p_free_solution}
    \\
    p_2(t)=p_2(0)-g_2t\,p(0)+\frac{g_1g_2t^2}{2}p_1(0).
    \label{p2_free_solution}
\end{gather}
The position equations give
\begin{equation}
    x_2(t)=x_2(0)+\frac{t}{m_2}p_2(0)-\frac{g_2t^2}{2m_2}p(0)+\frac{g_1g_2t^3}{6m_2}p_1(0),
    \label{x2_free_solution}
\end{equation}
and
\begin{equation}
\begin{split}
    x(t)= \,&x(0)+\left(\frac{t}{m}-\frac{g_2^2t^3}{6m_2}\right)p(0)+g_2t\,x_2(0)\\
    &
    +\left(-\frac{g_1t^2}{2m}+\frac{g_1g_2^2t^4}{24m_2}\right)p_1(0)+\frac{g_2t^2}{2m_2}p_2(0).
    \label{x_free_solution}
\end{split}
\end{equation}
The free-particle dynamics therefore exhibits polynomial rather than exponential growth.

Using the idea developed in Sec.~\ref{reduced_wigner_dynamics}, the phase space evolution of the system can be written as 
\begin{equation}
    R_S(t)=A_{\mathrm{fp}}(t)R_S(0)+B_{\mathrm{fp}}(t)R_E(0),
    \label{free_block_evolution}
\end{equation}
where
\begin{equation}
    A_{\mathrm{fp}}(t)=\begin{pmatrix}
        1&
        \frac{t}{m}-
        \frac{g_2^2t^3}{6m_2}
        \\[2mm]
        0&1
    \end{pmatrix},
    \label{A_free}
\end{equation}
and hence
\begin{equation}
    A_{\mathrm{fp}}^{-1}(t)  =  \begin{pmatrix}
        1&
        -\frac{t}{m}+ \frac{g_2^2t^3}{6m_2}
        \\[2mm]
        0&1
    \end{pmatrix}.
    \label{Ainv_free}
\end{equation}
In particular,
\begin{equation}
    A_{\mathrm{fp}}^{-1}(t) \not\rightarrow0.
    \label{Ainv_free_not_zero}
\end{equation}
Thus, the flattening mechanism differs from that we will encounter later in the harmonic oscillator case.

\subsection{Polynomial growth of the Weyl displacement}
\label{free_weyl_growth}
The reduced Weyl characteristic function of the system, $\chi_S$, is obtained by setting the environment displacement arguments to zero.
\begin{equation}
    \Xi_S = \begin{pmatrix}
        \xi_x&\xi_p&0&0&0&0
    \end{pmatrix}^{T},
    \label{system_weyl_vector_free}
\end{equation}
For quadratic evolution,
\begin{equation}
    \chi_S(\xi_x,\xi_p;t)=\chi_0\left[S_{\mathrm{fp}}^{-1}(t)\Xi_S\right].
    \label{chi_reduced_free}
\end{equation}

Using the exact free-particle evolution,
\begin{equation}
    \Xi(t)=S_{\mathrm{fp}}^{-1}(t)\Xi_S=\begin{pmatrix}
	\xi_x + \left(-\frac{t}{m}+\frac{g_2^2t^3}{6m_2}\right)\, \xi_p 
	\\[2mm]
        \xi_p
        \\[2mm]
        -g_1t\,\xi_x+\left(\dfrac{g_1t^2}{2m}-\dfrac{g_1g_2^2t^4}{24m_2}\right)\xi_p
        \\[2mm]
        -\dfrac{g_2t^2}{2m_2}\xi_p
        \\[2mm]
        0
        \\[2mm]
        g_2t\,\xi_p
    \end{pmatrix}.
    \label{etaE_free}
\end{equation}
Thus, when the coupling is nonzero,
\begin{equation}
    \xi_S\neq0
    \quad\Longrightarrow\quad
    \|\Xi(t)\| \longrightarrow\infty.
    \label{etaE_divergence_free}
\end{equation}
This is the origin of the shear-induced behavior. A non-zero Weyl displacement $\xi_S$ produces Weyl components whose norm diverges at long times. Thus, for any physical trace-class quantum state, the Weyl characteristic function is continuous and vanishes at infinity~\cite{Werner1984}, as
shown in Appendix~\ref{weyl_vanish_infinity}.

The zero component is fixed by normalization,
\begin{equation}
    \chi_S(0,t) = \on{Tr}\rho_S(t) = 1.
    \label{chiS_zero_mode_free}
\end{equation}
Thus,
\begin{equation}
    \chi_S(\xi_S,t)
    \longrightarrow
    \begin{cases}
        1,
        & \xi_S=0,
        \\[1mm]
        0,
        & \xi_S\neq0.
    \end{cases}
    \label{free_characteristic_limit}
\end{equation}

Eq.~\eqref{free_characteristic_limit} is the continuum result. It states that for non-zero $\xi_S$, the Weyl characteristic function vanishes asymptotically, whereas at $\xi_S = 0$, it is equal to $1$ on account of the normalization.

\subsection{Relation to finite-volume construction}
\label{free_uniform_regulated}

Following Sec.~\ref{discrete_Weyl_discussions}, we now give a finite-dimensional interpretation of the continuum result. At fixed $L$ and $N$, the finite Weyl basis contains one identity mode $D_{0,0}$ and a finite set of nonidentity modes $D_{r,s}$. The continuum result
\begin{equation}
    \chi_S(\xi_S,t)\longrightarrow0,
    \qquad
    \xi_S\neq0,
\end{equation}
shows that asymptotically all nonzero subsystem Weyl components are suppressed. In finite Weyl representation, if
\begin{equation}
    \chi_{0,0}=1,
    \qquad
    \chi_{r,s}=0,
    \quad
    (r,s)\neq(0,0),
\end{equation}
from Eqs.~\eqref{density_weyl_expansion}-\eqref{discrete_wigner_fourier}, it leads to maximally mixed state and a uniform Wigner function.

Substituting these asymptotic coefficients into the operator expansion \eqref{density_weyl_expansion}, every term except the identity disappears \cite{Schwinger1960, Wootters1987, Vourdas2004}. Hence the finite-dimensional analogue of the continuum asymptotic state is
\begin{equation}
    \rho_{\rm asym}^{(L,N)}=\frac{1}{d}\chi_{0,0}D_{0,0}^{\dagger}=\frac{\mathbf 1_{L,N}}{d}.
\end{equation}
Substituting $d=2N+1$, we get
\begin{equation}
    \rho_{\rm asym}^{(L,N)} = \frac{\mathbf 1_{L,N}}{2N+1}.
    \label{free_maximally_mixed}
\end{equation}
Thus, the regulated analogue of the continuum reduced state approaches the maximally mixed state.

Using Eq.~\eqref{discrete_wigner_fourier} together with the asymptotic coefficients in Eq.~\eqref{free_characteristic_limit}, all terms with $(r,s)\neq(0,0)$ vanish,
\begin{equation}
    W_{a,b,\,\textrm{asym}}^{(L,N)}(t)=\frac{1}{d^2} \chi_{0,0}=\frac{1}{d^2},
    \label{discrete_wigner_uniform}
\end{equation}
for every phase-space point $(a,b)$. Thus,
\begin{equation}
    W_{a,b\,\textrm{asym}}^{(L,N)}(t) \longrightarrow \frac{1}{(2N+1)^2}.
    \label{free_uniform_discrete_wigner}
\end{equation}
The limiting Wigner function is therefore independent of the phase-space coordinates.

To conclude, the asymptotic decoherence obtained in \cite{Adarsh:2026btz} admits the direct phase-space interpretation that the free-particle dynamics sends every nonzero argument of the Weyl characteristic function to infinity in the full phase-space under the inverse phase-space evolution. Consequently, the Weyl characteristic function tends to zero at every nonzero argument. In the regularized picture, the decay of all $\chi_{r,s}$ for $(r,s)\ne (0,0)$ implies a uniform Wigner function and a maximally mixed reduced state.

\section{Harmonic oscillator model: the hyperbolic mechanism}
\label{harmonic_riccati}

We now consider the harmonic oscillator model by restoring the confining potential. This is the important extension beyond \cite{Adarsh:2026btz}. Unlike the free-particle dynamics, the spectral structure of the harmonic oscillator model is closely connected to the coupling strength. We will find that above a critical value of the coupling, the quadratic flow enters a hyperbolic regime containing exponentially expanding and contracting directions. This provides a qualitatively different route to Wigner flattening, characterized by the asymptotic decay of the matrix $A^{-1}(t)$. Rather than treating the intrinsic dynamics perturbatively, we retain the full Hamiltonian, and obtain the solution in the symmetric sector.

The Hamiltonian is
\begin{equation}
\begin{split}
    H_{\mathrm{ho}}=&\frac{p^2}{2m}+\frac{p_1^2}{2m_1}+\frac{p_2^2}{2m_2}\\
    &+\frac{1}{2}m\omega^2x^2+\frac{1}{2}m_1\omega_1^2x_1^2+\frac{1}{2}m_2\omega_2^2x_2^2\\
    &+g_1xp_1+g_2px_2.
    \label{harmonic_hamiltonian}
\end{split}
\end{equation}
In the following exact solution, for the purpose of tractability, we restrict to
\begin{equation}
    m_1=m_2=m,
    \quad
    \omega_1=\omega_2=\omega,
    \quad
    g_1=g_2=g,
\end{equation}
with $m>0$, $\omega>0$, and $g>0$.

\subsection{Riccati formulation}
\label{harmonic_riccati_formulation}

As before, we write the symplectic matrix in system--environment block form,
\begin{equation}
    S(t)=\begin{pmatrix}
        A(t) & B(t)\\
        C(t) & D(t)
    \end{pmatrix}.
\end{equation}
The evolution equation $\dot S=MS$ gives
\begin{align}
    \dot A&=M_{11}A+M_{12}C,
    \\
    \dot C&=M_{21}A+M_{22}C.
\end{align}
Assuming that $A(t)$ is invertible on the interval under consideration, we
introduce
\begin{equation}
    Y(t) = C(t)A^{-1}(t).
    \label{Y_definition_harmonic}
\end{equation}
Differentiating this expression gives the matrix Riccati equation
\begin{equation}
    \dot Y=M_{21}+M_{22}Y-YM_{11}-YM_{12}Y.
    \label{Y_riccati_harmonic}
\end{equation}
At the same time,
\begin{equation}
    \dot A^{-1} = -A^{-1}K(t),
    \qquad
    K(t) = M_{11}+M_{12}Y(t).
    \label{Ainv_K_harmonic}
\end{equation}
The nonlinear dynamics is entirely contained in $Y(t)$. Once $Y(t)$ is
known, $A^{-1}(t)$ follows from the linear equation
\eqref{Ainv_K_harmonic}.

The matrix $Y(t)$ is $4\times2$, and we write
\begin{equation}
    Y = \begin{pmatrix}
        y_{11} & y_{12}\\
        y_{21} & y_{22}\\
        y_{31} & y_{32}\\
        y_{41} & y_{42}
    \end{pmatrix}.
    \label{Y_components_harmonic}
\end{equation}
Thus, the Riccati equation consists of eight coupled nonlinear equations. Their explicit form and reduction are discussed in the Appendix~\ref{full_riccati}.

The initial condition follows from $S(0)=I_6$, which gives $A(0)=I_2$, and $C(0)=0$, and hence $Y(0)=0$.

The physical trajectory lies on the invariant manifold
\begin{equation}
\begin{gathered}
    y_{42}=-y_{11},
    \quad
    y_{32}=-y_{21},\\
    y_{41}=m^2\omega^2y_{12},
    \quad
    y_{31}=m^2\omega^2y_{22}.
    \label{harmonic_invariant_manifold}
\end{gathered}
\end{equation}
This reduces the eight-dimensional Riccati system to a four-dimensional closed system. The full component equations and the proof of the invariance of this manifold are given in Appendix~\ref{riccati_reduction}.

To expose the structure of the reduced dynamics, we introduce
\begin{equation}
    u=y_{11}-y_{21},
    \qquad
    v=y_{12}-y_{22},
\end{equation}
and define the complex variables
\begin{equation}
    z=u+im\omega v,
    \qquad
    Z=y_{21}+im\omega y_{22}.
    \label{zZ_definition}
\end{equation}
These combinations reduce the four real Riccati equations exactly to
\begin{gather}
    \dot z = g(1-Zz),
    \label{z_complex_equation}
    \\
    \dot Z = -gZ^2-2i\omega Z-i\omega z,
    \label{Z_complex_equation}
\end{gather}
with $z(0)=Z(0)=0$.

The remaining nonlinear system can itself be linearized. Eliminating $z$ and introducing
\begin{equation}
    Z(t)=\frac{\dot{\mc W}(t)}{g\mc W(t)},
    \label{Z_W_substitution}
\end{equation}
gives the linear third-order equation
\begin{equation}
    \dddot{\mc W}+2i\omega\ddot{\mc W}+ig^2\omega\mc W=0.
    \label{W_third_order}
\end{equation}
The corresponding initial conditions are
\begin{equation}
    \mc W(0)=1,
    \qquad
    \dot{\mc W}(0)=0,
    \qquad
    \ddot{\mc W}(0)=0.
    \label{W_initial_conditions}
\end{equation}
Thus, the nonlinear Riccati dynamics relevant to the physical trajectory is encoded in a single linear differential equation. The detailed reduction from the component Riccati equations to Eq.~\eqref{W_third_order} is given in Appendix~\ref{linearization_spectral_structure}. It is to be noted that the substitution in Eq.~\ref{Z_W_substitution} is well-defined, when $\mathcal W$ is nonzero. In Appendix~\ref{nonvanishing_W}, we show that the $\mathcal W$ is non-vanishing for all finite times. 

\subsection{Spectral threshold}
\label{characteristic_threshold}

We seek solutions of Eq.~\eqref{W_third_order} of the form
\begin{equation}
    \mc W(t)=e^{rt}.
\end{equation}
The characteristic equation is
\begin{equation}
    r^3+2i\omega r^2+ig^2\omega=0.
    \label{r_characteristic}
\end{equation}
Writing $r=-i\lambda$, this becomes the real cubic
\begin{equation}
    \lambda^3 - 2\omega\lambda^2 + g^2\omega = 0.
    \label{lambda_cubic}
\end{equation}
Its discriminant is
\begin{equation}
    \Delta = g^2\omega^2 \left( 32\omega^2-27g^2 \right).
    \label{cubic_discriminant}
\end{equation}
The root structure, therefore, changes at
\begin{equation}
    g_c = \sqrt{\frac{32}{27}}\,\omega.
    \label{critical_coupling}
\end{equation}

For $g<g_c$, all three roots $\lambda_j$ are real, and the corresponding
characteristic exponents $r_j=-i\lambda_j$ are purely imaginary. Thus, the characteristic modes are oscillatory, with no exponential growth or decay. At $g=g_c$, two roots coincide, while the characteristic exponents still have vanishing real parts. The hyperbolic regime appears only for $g>g_c$. 

In this regime, the cubic has one real root and one complex-conjugate pair,
\begin{equation}
    \lambda_0\in\mathbb R,
    \qquad
    \lambda_+=\alpha+i\beta,
    \qquad
    \lambda_-=\alpha-i\beta,
    \qquad
    \beta>0.
    \label{lambda_complex_pair}
\end{equation}
The associated characteristic exponents are
\begin{equation}
    r_0=-i\lambda_0,
    \qquad
    r_+=\beta-i\alpha,
    \qquad
    r_-=-\beta-i\alpha.
    \label{r_complex_pair}
\end{equation}
Thus, the spectrum therefore contains two expanding directions and two contracting directions with rate $\beta$ and $-\beta$ respectively, together with an oscillatory pair. This expanding--contracting pair is the origin of the \textit{hyperbolic mechanism}.

The same threshold also marks a qualitative change in the spectrum of the full Hamiltonian generator $M$. To distinguish its eigenvalues from the roots $\lambda_j$ of Eq.~\ref{lambda_cubic}, we denote the eigenvalues of $M$ by $\mu$. Direct evaluation of the characteristic polynomial gives
\begin{equation}
    \det(\mu I_6-M)=(\mu^2+\omega^2)^3-2g^2\omega^2(\mu^2+\omega^2)+g^4\omega^2.
    \label{full_M_characteristic}
\end{equation}
Introducing $y=\mu^2+\omega^2$, the characteristic equation becomes
\begin{equation}
    y^3-2g^2\omega^2y+g^4\omega^2=0.
    \label{full_M_cubic}
\end{equation}Its discriminant is
\begin{equation}
    \Delta_M = g^6\omega^4 \left(32\omega^2-27g^2\right),
    \label{full_M_discriminant}
\end{equation}
so that the full generator undergoes its spectral transition at the same critical coupling, $g_c^2=\frac{32}{27}\omega^2$.

Thus, the critical coupling obtained from the Riccati equation is a manifestation of the genuine spectral transition of the full Hamiltonian generator $M$. The detailed structure of roots is given in Appendix~\ref{spectrum_of_M}, and illustrated in Fig.~\ref{spectral_transition}.
\begin{figure}[t]
    \centering
    \includegraphics[width = 0.5\textwidth,keepaspectratio]{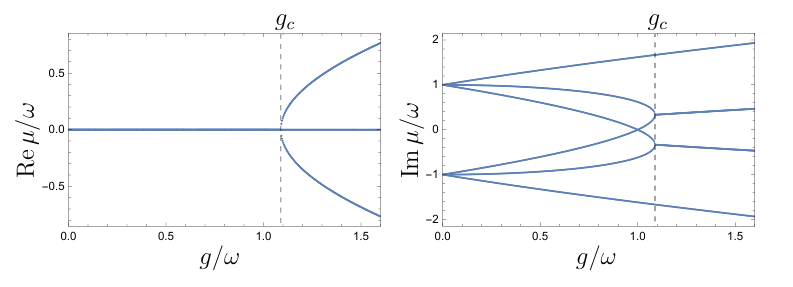}
    \caption{The real and imaginary parts of the eigenvalues of the Hamiltonian generator $M$, as functions of dimensionless parameter $g/\omega$. The dashed vertical line represents the critical coupling $g_c = \sqrt{32/27}\,\omega$. At $g<g_c$, the spectrum is purely imaginary. At $g=g_c$, two pairs of imaginary eigenvalues coalesce. At $g>g_c$, they emerge as a complex quartet with nonzero real parts. This clearly shows that the $g>g_c$ marks the hyperbolic regime.}
    \label{spectral_transition}
\end{figure}

\subsection{Asymptotic Riccati fixed point}
\label{asymptotic_Riccati_fixed_point}

For $g>g_c$, the characteristic equation contains an exponent with positive real part. For the physical initial conditions in Eq.~\eqref{W_initial_conditions}, the coefficient of this growing mode is nonzero, as shown in Appendix~\ref{physical_mode_selection}. Consequently, at long times,
\begin{equation}
    \mc W(t) = c_2\,e^{(\beta-i\alpha)t} \left[ 1+O(e^{-\beta t}) \right], 
    \qquad
    t\rightarrow\infty.
    \label{W_asymptotic}
\end{equation}
It follows that
\begin{equation}
    \frac{\dot{\mc W}(t)} {\mc W(t)} = \beta-i\alpha + O(e^{-\beta t}).
    \label{logW_asymptotic}
\end{equation}
Using Eq.~\eqref{Z_W_substitution}, we therefore obtain 
\begin{equation}
    Z(t) \longrightarrow Z_* = \frac{\beta-i\alpha}{g}.
    \label{Z_fixed_point}
\end{equation}
Since $Z=y_{21}+im\omega y_{22}$, the asymptotic values are
\begin{equation}
    y_{21}^{*} = \frac{\beta}{g}, 
    \qquad
    y_{22}^{*} = -\frac{\alpha}{gm\omega}.
    \label{y_fixed_points}
\end{equation}
In particular,
\begin{equation}
    gy_{21}^{*} = \beta > 0.
    \label{gy21_positive}
\end{equation}
This positive asymptotic value determines the exponential decay rate of $A^{-1}(t)$, as shown in the following subsection.

\subsection{Spectral solution for $A^{-1}(t)$}
\label{Ainv_exact_solution}

We return to Eq.~\eqref{Ainv_K_harmonic}. Using the invariant manifold relations in Eq.~\eqref{harmonic_invariant_manifold}, and special form of matrices $M_{12}$ and $M_{21}$ from Eqs.~\eqref{M12_appendix}--\eqref{M21_appendix} we obtain
\begin{equation}
    K(t) = \begin{pmatrix}
        gy_{21} & \dfrac{1}{m}+gy_{22}
        \\[2mm]
        -m\omega^2-gm^2\omega^2y_{22} & gy_{21}
    \end{pmatrix}.
    \label{K_invariant}
\end{equation}
This can be written as
\begin{equation}
    K(t)
    =
    a(t)I_2+b(t)J,
    \label{K_aIbJ}
\end{equation}
where $a(t)=gy_{21}(t)$, $b(t)=\frac{1}{m}+gy_{22}(t)$, and $J= \begin{pmatrix} 0&1\\ -m^2\omega^2&0 \end{pmatrix}$, $J^2=-m^2\omega^2I_2$. Since all time dependence is contained in the scalar functions $a(t)$ and
$b(t)$,
\begin{equation}
    [K(t_1),K(t_2)]=0
    \qquad
    \text{for all }t_1,t_2.
    \label{K_commutativity}
\end{equation}
The solution, therefore, requires no time ordering.

The eigenvalues of $K(t)$ are
\begin{equation}
    \kappa_{\pm}(t) = gy_{21}(t) \pm i\omega \left[ 1+gmy_{22}(t) \right].
    \label{K_eigenvalues}
\end{equation}
Because $K(t)$ is a linear combination of $I_2$ and the fixed matrix $J$, all $K(t)$ share the same eigenspaces. Their spectral projectors are therefore time independent; we denote them by $P_{\pm}$. Hence,
\begin{equation}
    K(t) = \kappa_+(t)P_+ + \kappa_-P_-.
\end{equation}
Thus,
\begin{equation}
\begin{split}
    A^{-1}(t) &=\exp\left[-\int_0^t K(s)ds\right]\\
    &= \sum_{\sigma=\pm} \exp \left[ -\int_0^t \kappa_{\sigma}(s)\,ds \right] P_{\sigma}.
    \label{Ainv_spectral_solution}
\end{split}
\end{equation}
Since the two eigenvalues have the same real part, $\on{Re}\kappa_{\pm}(t) = gy_{21}(t)$, the common decay envelope of the two spectral components of $A^{-1}(t)$ is
\begin{equation}
    \exp \left[ -g\int_0^t y_{21}(s)\,ds \right].
    \label{Ainv_envelope}
\end{equation}
The explicit spectral projectors are given in Appendix~\ref{spectral_Ainverse}.

\subsection{Hyperbolic Wigner flattening}
\label{harmonic_Wigner_flattening}

\begin{figure}[t]
    \centering
    \includegraphics[width = 0.5\textwidth,keepaspectratio]{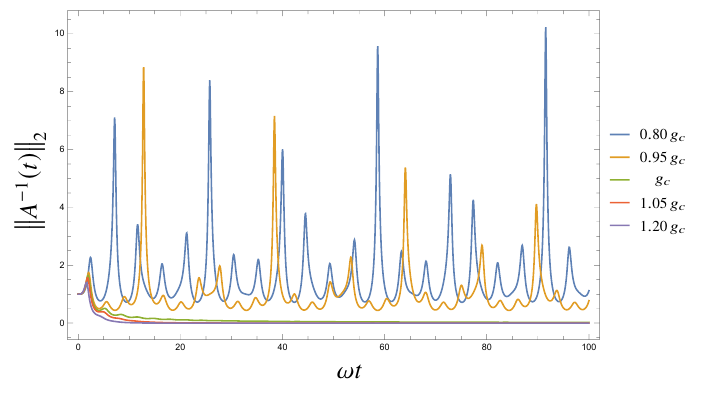}
    \caption{Time evolution of the spectral norm of $A^{-1}(t)$, at different values of the coupling strength relative to the critical coupling $g_c$. At $g<g_c$, it remains oscillatory and does not approach zero. While at $g>g_c$, $\|A^{-1}(t)\|_{2}$ decays towards zero, which illustrates the mechanism by which nonzero subsystem phase-space separations are suppressed to zero.}
    \label{Ainverse_norm}
\end{figure}

For $g>g_c$, Eq.~\eqref{gy21_positive} gives $gy_{21}^{*}=\beta>0$. Therefore,
\begin{equation}
    g\int_0^t y_{21}(s)\,ds = \beta t+O(1),
    \quad\text{as }
    t\rightarrow\infty,
\end{equation}
and the envelope in Eq.~\eqref{Ainv_envelope} decays exponentially. Hence,
\begin{equation}
    A^{-1}(t) = O(e^{-\beta t}) \longrightarrow0,
    \qquad
    g>g_c.
    \label{Ainv_harmonic_zero}
\end{equation}
The qualitative difference in the behavior of $A^{-1}(t)$ is illustrated in Fig.~\ref{Ainverse_norm}. The $\|A^{-1}(t)\|_2$, where $\|\cdot\|$ implies the $L^2$ norm, is plotted against time for various values of the coupling strength relative to the critical coupling. Below the threshold, $\|A^{-1}(t)\|_2$ does not decay, whereas above the threshold, it approaches zero, in agreement with the solution we found.

We now turn to the reduced Wigner function in Eq.~\eqref{reduced_Wigner_general}. For two subsystem phase-space points $R_S$ and $R'_S$,
\begin{widetext}
\begin{equation}
    W_S(R_S,t)-W_S(R'_S,t)=\frac{1}{|\det A(t)|}\int d^4u\,\Big[W_0\!\left(A^{-1}(t)R_S-F(t)u,u\right) -W_0\!\left( A^{-1}(t)R'_S-F(t)u,u\right)\Big].
    \label{Wigner_difference_harmonic}
\end{equation}
\end{widetext}
The two subsystem arguments of the initial Wigner function differ by
\begin{equation}
    A^{-1}(t)(R_S-R'_S).
    \label{contracted_Wigner_separation}
\end{equation}
Thus, for every fixed pair $R_S, R'_S$,
\begin{equation}
    \left\|A^{-1}(t)(R_S-R'_S)\right\|\longrightarrow0.
\end{equation}

For a compact region $K$, we have
\begin{equation}
    \|R_S-R'_S\| \leq \on{diam}(K),
    \qquad
    R_S,R'_S\in K,
\end{equation}
and hence,
\begin{equation}
    \sup_{R_S,R'_S\in K} \left\| A^{-1}(t)(R_S-R'_S) \right\| \leq \|A^{-1}(t)\| \on{diam}(K) \longrightarrow0.
    \label{uniform_contraction_K}
\end{equation}
Therefore, if the initial Wigner function is uniformly continuous in its subsystem argument, with sufficient uniformity and integrability in the environmental variables to pass the limit through the integral, the difference in Eq.~\eqref{Wigner_difference_harmonic} vanishes uniformly for $R_S,R'_S\in K$. The detailed analysis of the regularity requirement of the Wigner function can be found in Appendix~\ref{regularity_wigner_flattening}. Consequently,
\begin{equation}
    \Delta_K(t)=\sup_{R_S,R'_S\in K}\left|W_S(R_S,t)-W_S(R'_S,t)\right|\longrightarrow0.
    \label{harmonic_local_flattening_result}
\end{equation}

The role of $A^{-1}(t)$ is therefore direct: two distinct subsystem points are mapped to subsystem arguments whose separation is $A^{-1}(t)(R_S-R'_S)$, and this separation contracts exponentially in the hyperbolic regime $g>g_c$. The reduced Wigner function therefore loses its dependence on the subsystem phase-space coordinate over every fixed compact region. This is the hyperbolic route to local Wigner flattening.

\subsection{Maximally mixed state in the regularized harmonic model}
\label{harmonic_uniform_regularized}

The finite-volume regularization introduced in Sec.~\ref{finite_volume_regularization} provides a precise interpretation of the local Wigner flattening obtained for the harmonic oscillator model. The logic is slightly different from that of the free-particle case. There, the asymptotic behavior followed from the direct suppression of all nonzero subsystem Weyl components. In the harmonic case, the natural starting point is instead the contraction of subsystem phase-space separations generated by $A^{-1}(t)$.

For fixed $L$ and $N$, the regulated subsystem phase space contains a finite set of points,
\begin{equation}
    (a,b)\in\mathbb Z_d\times\mathbb Z_d,
    \qquad
    d=2N+1.
\end{equation}
Let $W_{a,b}^{(L,N)}(t)$ denote the corresponding discrete Wigner function. Since $L$ and $N$ are fixed, these phase-space points sample a fixed bounded region $K_{L,N}$ of the continuum phase space. The continuum flattening result, therefore, implies that the difference between the Wigner function evaluated at any two such regulated phase-space points vanishes at late times. Hence,
\begin{equation}
    \max_{a,b,a',b'}\left|W_{a,b}^{(L,N)}(t)-W_{a',b'}^{(L,N)}(t)\right|\longrightarrow0.
    \label{discrete_Wigner_difference_zero}
\end{equation}
Thus, at fixed $L$ and $N$, all values of the regulated Wigner function approach a common asymptotic value, which we denote by $C_{L,N}$.

The discrete Wigner function is normalized according to
\begin{equation}
    \sum_{a,b\in\mathbb Z_d}W_{a,b}^{(L,N)}(t)=1.
    \label{discrete_Wigner_normalization}
\end{equation}
Since the regulated phase space contains $d^2$ points, asymptotic uniformity gives $d^2 C_{L,N}=1$. Therefore,
\begin{equation}
    W_{a,b}^{(L,N)}(t)\longrightarrow\frac{1}{d^2}=\frac{1}{(2N+1)^2}.
    \label{harmonic_discrete_Wigner_uniform}
\end{equation}

The same conclusion is obtained in the Weyl representation. The discrete Wigner function and the Weyl characteristic coefficients are related by the finite symplectic Fourier transform in Eq.~\eqref{discrete_wigner_fourier}. A constant discrete Wigner function has only the zero Fourier component. Hence,
\begin{equation}
    \chi_{r,s}(t)\longrightarrow
    \begin{cases}
        1, &(r,s)=(0,0), \\[1mm]
        0,&(r,s)\neq(0,0).
    \end{cases}
    \label{harmonic_discrete_Weyl_limit}
\end{equation}
Using the finite Weyl expansion of the density operator,
\begin{equation}
    \rho_S^{(L,N)}(t)=\frac{1}{d}\sum_{r,s\in\mathbb Z_d}\chi_{r,s}(t)D_{r,s}^{\dagger},
\end{equation}
only the identity component survives asymptotically. Therefore,
\begin{equation}
    \rho_S^{(L,N)}(t) \longrightarrow \frac{\mathbf 1_{L,N}}{d} = \frac{\mathbf 1_{L,N}}{2N+1}.
    \label{harmonic_maximally_mixed}
\end{equation}
Thus, the finite-dimensional Weyl--Wigner representation of the asymptotically flattened continuum state is the maximally mixed state
\begin{equation}
    \rho_{\rm asym}^{(L,N)} = \frac{\mathbf 1_{L,N}}{2N+1}.
\end{equation}

As in the free-particle case, the finite-dimensional construction is used to interpret the asymptotic continuum behavior rather than to replace the continuous-variable dynamics. The finite-dimensional construction is applied only after the continuum long-time behavior has been established. Removing the regulators then returns to the noncompact continuum setting, where no normalizable maximally mixed state exists and the appropriate statement remains local Wigner flattening.

\subsection{Energetic Indefiniteness and Nonlinear Stabilization}
\label{energetic_indefiniteness}

Before we end this section, it is helpful to clarify the energetic structure of the harmonic model. In the symmetric sector ($m_1 = m_2 = m$, $\omega_1 = \omega_2 = \omega$, and $g_1 = g_2 = g$), completing the squares allows the Hamiltonian to be expressed as
\begin{equation}
\begin{split}
H_{\rm ho}&=\frac{(p_1+mgx)^2}{2m}+\frac{(p+mgx_2)^2}{2m}+\frac{p_2^2}{2m}+\frac{m\omega^2x_1^2}{2}\\
&\quad
+\frac{m}{2}(\omega^2-g^2)\left(x^2+x_2^2\right).
\label{Hho_completed_square}
\end{split}
\end{equation}
Consequently, the quadratic Hamiltonian is positive definite if and only if
\begin{equation}
g^2 < \omega^2.
\label{positive_energy_condition}
\end{equation}
By contrast, the hyperbolic regime derived below emerges at
\begin{equation}
g_c^2 = \frac{32}{27}\omega^2 > \omega^2.
\end{equation}
This confirms that the hyperbolic phase of the autonomous quadratic model lies entirely within an energetically indefinite sector~\cite{Kustura2019}. Notice that there is however an intermediate regime $\omega^2<g^2<g_c^2$ in which the quadratic Hamiltonian is already energetically indefinite, while the spectrum remains nonhyperbolic. Energetic indefiniteness alone is thus not sufficient to produce the hyperbolic flattening mechanism.

Here again we provide two constructions: firstly, by treating the quadratic Hamiltonian as an effective theory of another quartic theory around an unstable stationary point, and secondly, by confining the spatial coordinates in a finite box. Consider the nonlinear extension
\begin{equation}
    H_{\rm full}=H_{\rm ho}+\frac{\lambda}{4}\left(x^4+x_2^4\right),
    \qquad
    \lambda>0.
\label{nonlinear_parent}
\end{equation}
Using Eq.~\eqref{Hho_completed_square}, this becomes
\begin{equation}
\begin{split}
H_{\rm full}&=\frac{(p_1+mgx)^2}{2m}+\frac{(p+mgx_2)^2}{2m}+\frac{p_2^2}{2m}+\frac{m\omega^2x_1^2}{2}\\
&\quad
+\frac{m}{2}(\omega^2-g^2)\left(x^2+x_2^2\right)+\frac{\lambda}{4}\left(x^4+x_2^4\right).
\label{nonlinear_parent_completed}
\end{split}
\end{equation}
Because $\lambda > 0$, the quartic potential dominates at large amplitudes, guaranteeing that $H_{\rm full}$ is globally bounded from below. However, for $g^2 > \omega^2$, the origin becomes an unstable stationary point of the potential along the $(x, x_2)$ directions. Expanding $H_{\rm full}$ around this origin gives
\begin{equation}
H_{\rm full} = H_{\rm ho} + \mc{O}\left(x^4, x_2^4\right).
\end{equation}
Hence, $H_{\rm ho}$ serves as the exact quadratic effective Hamiltonian describing the local linearized dynamics sufficiently close to an unstable stationary point of a nonlinear parent system that is bounded from below~\cite{Barton1986,Subramanyan2021}.

A simple scaling estimate makes this validity window explicit. Along an unstable direction $q$, the local potential takes the schematic form
\begin{equation}
V(q) \simeq -\frac{a}{2}q^2 + \frac{\lambda_{\rm eff}}{4}q^4,     \qquad a > 0,
\end{equation}
where $a$ is of order $m(g^2 - \omega^2)$, up to numerical factors determined by the unstable eigenvector. The nonlinear term becomes comparable to the quadratic contribution at the characteristic amplitude
\begin{equation}
    q_{\rm nl} \sim \sqrt{\frac{2a}{\lambda_{\rm eff}}}.
\end{equation}
Given an initial displacement $q_0$ along the unstable manifold and a linearized growth rate $\beta$, the trajectory evolves as $q(t) \sim q_0 e^{\beta t}$, yielding the logarithmic crossover time
\begin{equation}
    t_{\rm nl} \sim \frac{1}{\beta} \log\!\left(\frac{q_{\rm nl}}{\vert{}q_0\vert{}}\right).
\end{equation}
Here, $\beta$ represents the exact positive growth rate obtained from the full three-mode spectral problem in Sec.~\ref{characteristic_threshold}. Consequently, the quadratic hyperbolic mechanism possesses a parametrically well-resolved validity window whenever $q_{\rm nl} \gg \vert q_0\vert$. This is an estimate for a classical amplitude envelope, not a bound on the quantum evolution error. For a wave packet, its spread along the expanding directions must also remain small compared with the nonlinear scale.

The exact analysis presented above applies directly to this quadratic effective theory. Consequently, the strict long-time limit derived within the quadratic model should not be interpreted as a proof of asymptotic flattening for the full nonlinear Hamiltonian in Eq.~\eqref{nonlinear_parent}. In the complete parent theory, the stabilizing quartic terms inevitably take effect as the phase-space trajectory departs from the neighborhood of the unstable fixed point. Instead, the quadratic result isolates the fundamental hyperbolic mechanism, providing an exact description on time scales where the linearized dynamics remains valid.

In the regularized setting (finite-volume and momentum cutoff) introduced in Sec.~\ref{finite_volume_regularization}, where the notion of maximally mixed state makes rigorous sense, the Hamiltonian is bounded from below. In Appendix~\ref{finite_volume_robustness}, we show that in a finite-volume setting, for every finite time interval, the finite-volume Wigner function can be shown to approach the continuum Wigner function by choosing a sufficiently large volume, showing that the mechanism does not necessarily require a Hamiltonian that is unbounded from below.

\section{Discussion and conclusion}
\label{discussion}

The present work extends the conjugate-measurement mechanism introduced in Ref.~\cite{Adarsh:2026btz}. In that work, we established that coherence in both conjugate bases is suppressed, yielding a uniform, reduced statistical description in the interaction-dominated regime. The present analysis shows how that behavior emerges from the exact symplectic geometry when the intrinsic quadratic Hamiltonians are retained.

The composite system considered in this work consists of one distinguished subsystem and two auxiliary environmental modes. The analysis follows the exact unitary dynamics without invoking a Markov approximation or a continuum reservoir \cite{Gorini1976, Lindblad1976}. Nevertheless, the reduced subsystem state exhibits a loss of local phase-space dependence. This shows that strong phase-space mixing between the subsystem and environment modes produces an equilibration-like reduced dynamics even in finite-mode closed systems \cite{GemmerMichel2006, GemmerMichel2006EPJB}.

The results of the earlier work are not restricted to Gaussian states. The present symplectic analysis makes this transparent. For quadratic Hamiltonians, the Weyl characteristic functions evolve by composition with $S^{-1}(t)$ as $\chi_t(\xi)=\chi_0\!\left(S^{-1}(t)\xi\right)$, where $S(t)$ is the symplectic matrix \cite{Hillery1984, MukundaSimonSudarshan1988}. The above equation holds irrespective of the functional form of the initial states. Consequently, the asymptotic dynamics derived here applies to a wide class of states, subject only to the regularity assumptions of the Wigner state explained in Appendix~\ref{regularity_wigner_flattening}. Related studies of quadratic dynamics in extended many-body settings have also shown that local relaxation need not require Gaussian initial states \cite{CramerEisert2010, MurthySrednicki2019}.

The reduced local flattening of the subsystem remains fully consistent with the global unitary evolution of the three-mode composite system. At the level of the full state, no information is destroyed. Instead, the information initially contained in the subsystem phase space is distributed to the environmental degrees of freedom via correlations between the subsystem and environmental modes. Thus, even though the full dynamics remains reversible, the reduced state becomes progressively insensitive to the subsystem coordinates. The apparent local flattening occurs by reducing the description to the subsystem in an otherwise information-preserving global dynamical state \cite{Linden2009, CramerDawsonEisertOsborne2008, CramerEisert2010}.
 
The harmonic oscillator model further provides a direct link between reduced-state flattening and unstable quadratic dynamics. In the hyperbolic regime $g>g_c$, the characteristic exponents acquire real parts, giving rise to an expanding-contracting pair. The corresponding symplectic flow exhibits exponentially growing and decaying phase-space directions \cite{Kustura2019, Subramanyan2021}. For the reduced system, the relevant dynamics is reflected in the decay of $A^{-1}(t)$, which maps distinct final subsystem phase-space points back to increasingly nearby subsystem arguments in the initial Wigner function. The resulting loss of local phase-space resolution is therefore geometric in origin.

Unstable quadratic Hamiltonians are known to generate entanglement growth governed by the unstable directions of the corresponding classical phase-space flow. In particular, these unstable sectors produce linear entanglement growth at rates dictated by the appropriate Lyapunov exponents of the underlying symplectic dynamics \cite{Hackl2018, DePalma2022}. Here, we identify a complementary reduced-state consequence of this same unstable phase-space geometry. The inverse subsystem block thereby contracts finite phase-space separations. This demonstrates that the underlying instability not only generates entanglement growth, but also drives a geometrically identifiable loss of local subsystem phase-space resolution.

Crucially, this phenomenon must not be identified with thermalization in the standard statistical-mechanical sense \cite{Reimann2008, Linden2009, GogolinEisert2016}. The dynamics establishes neither relaxation to a Gibbs state nor the selection of a temperature. In the continuum limit, the appropriate asymptotic description is local Wigner flattening rather than convergence to a globally uniform, normalizable state. As shown in Appendix~\ref{finite_volume_robustness}, the finite-volume estimate guarantees that the continuum flattening observed over finite times persists, within controlled error bounds, under sufficiently large spatial confinement.

These results demonstrate that equilibration-like reduced dynamics can emerge directly from the geometry of a finite-mode closed Hamiltonian flow. The present framework naturally points toward extensions involving larger environmental reservoirs. Notably, the two environmental modes can be viewed as the minimal finite-mode precursor to a continuum of harmonic oscillators, a setting that forms a standard class of system--reservoir models \cite{FeynmanVernon1963, MonnaiMorodomeYuasa2019}. This raises the question of whether the shear-induced and hyperbolic mechanisms persist when a continuous spectrum replaces the discrete environment, with a defined spectral density. Within such a continuous setting, it remains to be determined whether the local subsystem flattening holds generically under continuum couplings. Establishing this would provide a link between the exactly solvable three-mode dynamics presented here and more realistic open-quantum-system models.

\section{Acknowledgment}
The authors used ChatGPT (GPT-6, OpenAI) and Gemini 3.5 Flash (Google) to assist with coding, literature searches, and language editing. The analysis and scientific interpretations presented were performed entirely by the authors, who take full responsibility for the content and integrity of the manuscript.

\section*{Data Availability}
No data were created or analyzed in this study. The Mathematica code used to generate the figures is available from the authors upon reasonable request.

\appendix

\section{Exact reduced Wigner dynamics}
\label{reduced_wigner}

In this appendix, we derive the reduced Wigner formula used in Sec.~\ref{general_mechanism}. We begin from the exact symplectic evolution of the full Wigner function,
\begin{equation}
    W_t(R)=W_0\!\left(S^{-1}(t)R\right),
\end{equation}
where the phase-space coordinate is decomposed as
\begin{equation}
    R=
    \begin{pmatrix}
        R_S\\
        R_E
    \end{pmatrix},
    \qquad
    R_S\in\mathbb{R}^{2},
    \qquad
    R_E\in\mathbb{R}^{4}.
\end{equation}
The reduced Wigner function of the subsystem is obtained by integrating over the environment variables,
\begin{equation}
    W_S(R_S,t)=\int d^4R_E\,W_0\!\left[S^{-1}(t)
    \begin{pmatrix}
        R_S\\
        R_E
    \end{pmatrix}
    \right].
    \label{redwigner_start}
\end{equation}

We write the symplectic propagator and its inverse in the block form, under the subsystem--environment decomposition,
\begin{equation}
    S(t) = \begin{pmatrix}
        A & B\\
        C & D
    \end{pmatrix},
    \qquad
    S^{-1}(t) = \begin{pmatrix}
        E & F\\
        G & H
    \end{pmatrix},
    \label{block_forms}
\end{equation}
where $A,E$ are $2\times 2$, $B,F$ are $2\times 4$, $C,G$ are $4\times 2$ and $D,H$ are $4\times 4$. The argument of $W_0$ in Eq.~\eqref{redwigner_start} is therefore
\begin{equation}
    S^{-1}
    \begin{pmatrix}
        R_S\\
        R_E
    \end{pmatrix} = \begin{pmatrix}
        ER_S+FR_E\\
        GR_S+HR_E
    \end{pmatrix}.
    \label{inverse_action}
\end{equation}

We now introduce a new environmental integration variable
\begin{equation}
    u=GR_S+HR_E.
    \label{u_definition}
\end{equation}
For fixed $R_S$, and when $H$ is invertible,
\begin{equation}
    R_E = H^{-1}(u-GR_S),
\end{equation}
and hence
\begin{equation}
    d^4R_E = \frac{d^4u}{|\det H|}.
    \label{jacobian_H}
\end{equation}
The subsystem component in Eq.~\eqref{inverse_action} then becomes
\begin{equation}
\begin{split}
    ER_S+FR_E&=ER_S+FH^{-1}(u-GR_S)\\
    &=\left(E-FH^{-1}G\right)R_S+FH^{-1}u.
    \label{subsystem_argument_general}
\end{split}
\end{equation}
Equation~\eqref{redwigner_start}, therefore, takes the form
\begin{widetext}
\begin{equation}
    W_S(R_S,t)=\frac{1}{|\det H|}\int d^4u\,W_0\!\left(\left[E-FH^{-1}G\right]R_S+FH^{-1}u,u\right).
    \label{reduced_EFGH}
\end{equation}
\end{widetext}
We now want to express the combinations in Eq.~\eqref{reduced_EFGH} in terms of the blocks of $S(t)$. Assuming $A(t)$ is invertible (at the end of the solution, it becomes clear that for the given initial conditions, $A(t)$ is always invertible), the Schur complement
\begin{equation}
    K_A = D-CA^{-1}B
    \label{schur_A}
\end{equation}
is also invertible, because $\on{det}(A)\cdot\on{det}(K_A) = \on{det}(S) = 1$. 

The standard block-inverse formula gives
\begin{equation}
    S^{-1} = 
    \begin{pmatrix}
        A^{-1}+A^{-1}BK_A^{-1}CA^{-1}
        &
        -A^{-1}BK_A^{-1}
        \\[2mm]
        -K_A^{-1}CA^{-1}
        &
        K_A^{-1}
    \end{pmatrix}.
    \label{block_inverse_A}
\end{equation}
Comparing Eqs.~\eqref{block_forms} and \eqref{block_inverse_A}, we identify
\begin{equation}
    H=K_A^{-1},
    \qquad
    F=-A^{-1}BK_A^{-1}.
\end{equation}
It follows immediately that
\begin{equation}
    FH^{-1} = -A^{-1}B.
    \label{FH_relation}
\end{equation}

Using Eq.~\eqref{block_inverse_A},
\begin{align}
    E&=A^{-1}+A^{-1}BK_A^{-1}CA^{-1},\\
    G&=-K_A^{-1}CA^{-1}.
\end{align}
Therefore,
\begin{equation}
\begin{split}
    FH^{-1}G&=\left(-A^{-1}B\right)\left(-K_A^{-1}CA^{-1}\right)\\
    &= A^{-1}BK_A^{-1}CA^{-1},
\end{split}
\end{equation}
and hence
\begin{equation}
    E-FH^{-1}G = A^{-1}.
    \label{schur_identity}
\end{equation}
Thus the subsystem argument in Eq.~\eqref{subsystem_argument_general} reduces to
\begin{equation}
    A^{-1}R_S-A^{-1}Bu.
    \label{reduced_argument}
\end{equation}

Finally, we express the Jacobian in terms of $A$. From Eq.~\eqref{schur_A},
\begin{equation}
    H^{-1} = D-CA^{-1}B.
\end{equation}
The determinant identity for a block matrix with invertible $A$ gives
\begin{equation}
    \det S = \det A\, \det\!\left(D-CA^{-1}B\right).
    \label{block_det}
\end{equation}
Since $S(t)$ is symplectic,
\begin{equation}
    \det S(t)=1.
\end{equation}
Consequently,
\begin{equation}
    \det H^{-1} = \frac{1}{\det A},
\end{equation}
and therefore
\begin{equation}
    \frac{1}{|\det H|} = \frac{1}{|\det A|}.
    \label{jacobian_A}
\end{equation}

Substituting Eqs.~\eqref{FH_relation}, \eqref{schur_identity}, and \eqref{jacobian_A} into Eq.~\eqref{reduced_EFGH}, we obtain
\begin{widetext}
\begin{equation}
    W_S(R_S,t)=\frac{1}{|\det A(t)|}\int d^4u\,W_0\!\left(A^{-1}(t)R_S-A^{-1}(t)B(t)u,u\right).
    \label{exact_reduced_wigner}
\end{equation}
\end{widetext}

Defining
\begin{equation}
    F(t) \equiv A^{-1}(t)B(t),
\end{equation}
the same expression can be written as
\begin{equation}
    W_S(R_S,t)=\frac{1}{|\det A(t)|}\int d^4u\,W_0\!\left(A^{-1}(t)R_S-F(t)u,u\right).
    \label{exact_reduced_wigner_F}
\end{equation}
Equation~\eqref{exact_reduced_wigner_F} is the form used throughout the main text. It separates the explicit subsystem dependence, controlled by $A^{-1}(t)$, from the environmental contribution, controlled by $F(t)=A^{-1}(t)B(t)$.

\section{Vanishing of the Weyl characteristic function at infinity}
\label{weyl_vanish_infinity}

In the free-particle analysis in Sec.~\ref{free_particle_section} we use the fact that the Weyl characteristic function of a trace-class quantum state vanishes when the phase-space displacement becomes arbitrarily large. We give a short proof here.

We first consider a rank-one operator
\begin{equation}
    \rho_{\phi,\psi}=\ket{\psi}\bra{\phi},
\end{equation}
for which
\begin{equation}
    \chi_{\phi,\psi}(\xi) = \bra{\phi}D(\xi)\ket{\psi},
\end{equation}
where $\chi_{\phi,\psi}$ is the Weyl characteristic function corresponding to the operator $\rho_{\phi,\psi}$, and $D(\xi)$ is the Weyl displacement operator corresponding to the phase-space displacement $\xi$. For one mode, writing $\xi=(q,p)$, an irrelevant phase convention aside, this matrix element has the form
\begin{equation}
    \chi_{\phi,\psi}(q,p) = e^{i\vartheta(q,p)} \int_{\mathbb R}dx\, \phi^{*}(x)\psi(x+q)e^{-ipx/\hbar}.
    \label{rankone_weyl}
\end{equation}
Its modulus therefore, satisfies
\begin{equation}
    |\chi_{\phi,\psi}(q,p)| \leq \int_{\mathbb R}dx\, |\phi(x)|\,|\psi(x-q)|.
    \label{rankone_bound}
\end{equation}
The right-hand side is the convolution of two $L^{2}$ functions, and hence is continuous and vanishes at infinity. Therefore,
\begin{equation}
    |\chi_{\phi,\psi}(q,p)| \longrightarrow0
    \qquad
    \text{as}
    \qquad
    |q|\longrightarrow\infty .
    \label{q_vanish}
\end{equation}
This bound is independent of the variation in values of $p$, so the decay as $|q|\to\infty$ is uniform in $p$.

Now, for fixed $q$, define a function
\begin{equation}
    f_q(x) = \phi^{*}(x)\psi(x-q).
\end{equation}
By the Cauchy--Schwarz inequality,
\begin{equation}
    \|f_q\|_{1} \leq \|\phi\|_{2}\|\psi\|_{2},
\end{equation}
so that $f_q\in L^{1}(\mathbb R)$. Eqn.~\eqref{rankone_weyl} is therefore, apart from a phase, the Fourier transform of $f_q$. The Riemann--Lebesgue lemma gives
\begin{equation}
    \chi_{\phi,\psi}(q,p) \longrightarrow0
    \qquad
    \text{as}
    \qquad
    |p|\longrightarrow\infty
\end{equation}
for every fixed $q$. However, now we should be careful about the possibility that the value of $q$ may change as we take the limit $p\to\infty$. This is important since we are interested in the limit $q^2 + p^2 \to \infty$, and not just along certain directions. Hence, we have to show that the decay as $p\to\infty$ is uniform, while $q$ remains in a bounded interval.

For two values $q$ and $q'$, using Cauchy-Schwarz inequality,
\begin{equation}
\begin{split}
    \|f_q-f_{q'}\|_1&=\int dx\,|\phi(x)|\left|\psi(x-q)-\psi(x-q')\right|\\
    &\leq\|\phi\|_2\,\|\psi(\cdot-q)-\psi(\cdot-q')\|_2,
    \label{translation_continuity_bound}
\end{split}
\end{equation}
Since translations are continuous in $L^2(\mathbb R)$,
\begin{equation}
    \|\psi(\cdot-q)-\psi(\cdot-q')\|_2 \longrightarrow0
    \qquad
    \text{as }q'\to q.
\end{equation}
Therefore, the map $q\longmapsto f_q$ is continuous from any bounded interval of $q$ into $L^1(\mathbb R)$.

If $q$ is restricted to a compact interval, the corresponding set $\{f_q\}$ is therefore compact in $L^1(\mathbb R)$. The Riemann--Lebesgue decay is then uniform over this compact family. Hence, for every bounded set of values of $q$,
\begin{equation}
    \sup_{q} \left| \int dx\,f_q(x)e^{ipx/\hbar} \right| \longrightarrow0,
    \qquad
    |p|\to\infty.
    \label{uniform_RL_decay}
\end{equation}

Combining this with Eq.~\eqref{q_vanish}, we obtain
\begin{equation}
    \bra{\phi}D(q,p)\ket{\psi} \longrightarrow0 \quad
    \text{as }
    q^{2}+p^{2}\longrightarrow\infty .
    \label{rankone_C0}
\end{equation}
The same argument applies to an arbitrary finite number of canonical modes for each component. No factorization of the wavefunctions between modes is required.

We now consider a general trace-class operator. Assuming that the operator is normalized, using singular-value decomposition, it may be written as
\begin{equation}
\begin{gathered}
    \rho = \sum_{j=1}^{\infty} s_j \ket{\psi_j}\bra{\phi_j},
    \qquad
    s_j\geq0,\\
    \sum_{j=1}^{\infty}s_j = \|\rho\|_1< \infty ,
\end{gathered}
\end{equation}
where $\ket{\psi_j}$ and $\ket{\phi_j}$ can be chosen to be normalized. For a density operator, $\|\rho\|_1=1$. The Weyl characteristic function of the operator is
\begin{equation}
    \chi_{\rho}(\xi) = \sum_{j=1}^{\infty} s_j \bra{\phi_j}D(\xi)\ket{\psi_j}.
    \label{traceclass_sum}
\end{equation}
By Cauchy--Schwarz inequality,
\begin{equation}
    \left| \bra{\phi_j}D(\xi)\ket{\psi_j} \right| \leq \|\ket{\phi_j}\|\cdot\|D(\xi)\ket{\psi_j}\| = 1.
\end{equation}
The last line follows from the normalization of the states $\psi_j$ and $\phi_j$, and the unitarity of $D(\xi)$. Because $\sum_j s_j = 1$, the right-hand side of Eq.~\eqref{traceclass_sum} converges uniformly in $\xi$. Every term in the series is continuous and vanishes at infinity by Eq.~\eqref{rankone_C0}. Hence, a uniformly convergent sum of such functions has the same property. Hence
\begin{equation}
    \chi_{\rho}\in C_{0}(\mathbb R^{2n}),
\end{equation}
where $C_{0}(\mathbb R^{2n})$  represents the space of all continuous functions on $\mathbb{R}^{2n}$ that vanish at infinity. Hence,
\begin{equation}
    \lim_{\|\xi\|\to\infty} \chi_{\rho}(\xi)=0 .
    \label{weyl_C0_result}
\end{equation}

A density operator is always trace-class. Therefore, in the free-particle model, if a nonzero final subsystem displacement $\xi_S$ satisfies
\begin{equation}
    \left\| S^{-1}(t)
    \begin{pmatrix}
        \xi_S\\
        0
    \end{pmatrix}
    \right\| \longrightarrow\infty ,
\end{equation}
then Eq.~\eqref{weyl_C0_result} directly implies
\begin{equation}
    \chi_S(\xi_S,t)=\chi_0\!\left[S^{-1}(t)
        \begin{pmatrix}
            \xi_S\\
            0
        \end{pmatrix}
    \right]
    \longrightarrow0,
    \qquad
    \xi_S\neq0.
\end{equation}
Thus, the suppression of the nonzero subsystem Weyl modes follows from the trace-class property of the initial quantum state and does not require a Gaussian assumption.

\section{Harmonic generator and full Riccati equations}
\label{full_riccati}

In this appendix, we develop the full-component form of the Riccati system for $Y(t) = C(t)A^{-1}(t)$ introduced in the Sec.~\ref {harmonic_riccati_formulation}. The solution described in that section is obtained in the symmetric sector
\begin{equation}
    m_1=m_2=m,
    \qquad
    \omega_1=\omega_2=\omega,
    \qquad
    g_1=g_2=g.
    \label{symmetric_sector_appendix}
\end{equation}
The Hamiltonian is therefore
\begin{equation}
\begin{split}
    H&=\frac{p^2}{2m}+\frac{p_1^2}{2m}+\frac{p_2^2}{2m}+\frac{1}{2}m\omega^2x^2\\
    &\qquad+\frac{1}{2}m\omega^2x_1^2+\frac{1}{2}m\omega^2x_2^2\\
    &\qquad+gxp_1+gpx_2.
    \label{harmonic_hamiltonian_appendix}
\end{split}
\end{equation}

For the canonical ordering
\begin{equation}
    R=
    \begin{pmatrix}
        x & p & x_1 & x_2 & p_1 & p_2
    \end{pmatrix}^{T},
\end{equation}
the Heisenberg equations are of the form
\begin{equation}
    \dot R=MR,
\end{equation}
with
\begin{equation}
    M=
    \begin{pmatrix}
        0 & \dfrac{1}{m} & 0 & g & 0 & 0\\[2mm]
        -m\omega^2 & 0 & 0 & 0 & -g & 0\\[2mm]
        g & 0 & 0 & 0 & \dfrac{1}{m} & 0\\[2mm]
        0 & 0 & 0 & 0 & 0 & \dfrac{1}{m}\\[2mm]
        0 & 0 & -m\omega^2 & 0 & 0 & 0\\[2mm]
        0 & -g & 0 & -m\omega^2 & 0 & 0
    \end{pmatrix}.
    \label{full_harmonic_generator}
\end{equation}

For the subsystem--environment decomposition, the variables are conveniently grouped as
\begin{equation}
    R_S=
    \begin{pmatrix}
        x\\
        p
    \end{pmatrix},
    \qquad
    R_E=
    \begin{pmatrix}
        x_1\\
        x_2\\
        p_1\\
        p_2
    \end{pmatrix}.
\end{equation}

In this ordering, the generator is written as
\begin{equation}
    M=
    \begin{pmatrix}
        M_{11} & M_{12}\\
        M_{21} & M_{22}
    \end{pmatrix},
    \label{M_block_appendix}
\end{equation}
where
\begin{equation}
    M_{11} =
    \begin{pmatrix}
        0 & \dfrac{1}{m}\\[2mm]
        -m\omega^2 & 0
    \end{pmatrix},
    \label{M11_appendix}
\end{equation}
\begin{equation}
    M_{12} =
    \begin{pmatrix}
        0 & g & 0 & 0\\
        0 & 0 & -g & 0
    \end{pmatrix},
    \label{M12_appendix}
\end{equation}
\begin{equation}
    M_{21} =
    \begin{pmatrix}
        g & 0\\
        0 & 0\\
        0 & 0\\
        0 & -g
    \end{pmatrix},
    \label{M21_appendix}
\end{equation}
and
\begin{equation}
    M_{22} =
    \begin{pmatrix}
        0 & 0 & \dfrac{1}{m} & 0\\[2mm]
        0 & 0 & 0 & \dfrac{1}{m}\\[2mm]
        -m\omega^2 & 0 & 0 & 0\\[2mm]
        0 & -m\omega^2 & 0 & 0
    \end{pmatrix}.
    \label{M22_appendix}
\end{equation}

We similarly decompose the symplectic propagator as
\begin{equation}
    S(t) =
    \begin{pmatrix}
        A(t) & B(t)\\
        C(t) & D(t)
    \end{pmatrix}.
\end{equation}
The evolution equation
\begin{equation}
    \dot S=MS
\end{equation}
gives, in particular,
\begin{equation}
    \dot A=M_{11}A+M_{12}C,
    \qquad
    \dot C=M_{21}A+M_{22}C.
    \label{AC_equations_appendix}
\end{equation}
Assuming $A(t)$ is invertible on the interval under consideration, we introduce
\begin{equation}
    Y(t)=C(t)A^{-1}(t).
    \label{Y_definition_appendix}
\end{equation}
Differentiating,
\begin{equation}
    \dot Y = \dot C A^{-1} + C\dot A^{-1}.
\end{equation}
Using
\begin{equation}
    \dot A^{-1} = -A^{-1}\dot A A^{-1},
\end{equation}
together with Eq.~\eqref{AC_equations_appendix}, we obtain
\begin{equation}
\begin{split}
    \dot Y &= \left(M_{21}A+M_{22}C\right)A^{-1} - CA^{-1} \left(M_{11}A+M_{12}C\right)A^{-1}\\
    &= M_{21} + M_{22}Y - YM_{11} - YM_{12}Y.
\end{split}
\end{equation}
Hence,
\begin{equation}
    \dot Y = M_{21}+M_{22}Y-YM_{11}-YM_{12}Y.
    \label{full_riccati_matrix}
\end{equation}

The matrix $Y$ has dimension $4\times2$, and we write
\begin{equation}
    Y=
    \begin{pmatrix}
        y_{11} & y_{12}\\
        y_{21} & y_{22}\\
        y_{31} & y_{32}\\
        y_{41} & y_{42}
    \end{pmatrix}.
    \label{Y_components_appendix}
\end{equation}
The three matrix products on the right-hand side of Eq.~\eqref{full_riccati_matrix} can now be evaluated explicitly:
\begin{equation}
    M_{22}Y =
    \begin{pmatrix}
        \dfrac{y_{31}}{m} & \dfrac{y_{32}}{m}\\[2mm]
        \dfrac{y_{41}}{m} & \dfrac{y_{42}}{m}\\[2mm]
        -m\omega^2y_{11} & -m\omega^2y_{12}\\[2mm]
        -m\omega^2y_{21} & -m\omega^2y_{22}
    \end{pmatrix},
    \label{M22Y_appendix}
\end{equation}
\begin{equation}
    YM_{11} =
    \begin{pmatrix}
        -m\omega^2y_{12} & \dfrac{y_{11}}{m}\\[2mm]
        -m\omega^2y_{22} & \dfrac{y_{21}}{m}\\[2mm]
        -m\omega^2y_{32} & \dfrac{y_{31}}{m}\\[2mm]
        -m\omega^2y_{42} & \dfrac{y_{41}}{m}
    \end{pmatrix}.
    \label{YM11_appendix}
\end{equation}
\begin{equation}
    YM_{12} =
    \begin{pmatrix}
        0 & gy_{11} & -gy_{12} & 0\\
        0 & gy_{21} & -gy_{22} & 0\\
        0 & gy_{31} & -gy_{32} & 0\\
        0 & gy_{41} & -gy_{42} & 0
    \end{pmatrix},
\end{equation}
\begin{equation}
    YM_{12}Y = g
    \begin{pmatrix}
        y_{11}y_{21}-y_{12}y_{31}
        &
        y_{11}y_{22}-y_{12}y_{32}
        \\[1mm]
        y_{21}^2-y_{22}y_{31}
        &
        y_{21}y_{22}-y_{22}y_{32}
        \\[1mm]
        y_{31}y_{21}-y_{32}y_{31}
        &
        y_{31}y_{22}-y_{32}^2
        \\[1mm]
        y_{41}y_{21}-y_{42}y_{31}
        &
        y_{41}y_{22}-y_{42}y_{32}
    \end{pmatrix}.
    \label{YM12Y_appendix}
\end{equation}

Substituting Eqs.~\eqref{M22Y_appendix}--\eqref{YM12Y_appendix} into Eq.~\eqref{full_riccati_matrix} gives the full system of eight coupled nonlinear equations,
\begin{align}
    \dot y_{11} &= \frac{y_{31}}{m} + m\omega^2y_{12} - g\left(y_{11}y_{21}-y_{12}y_{31}\right) + g,
    \label{riccati_y11}
    \\
    \dot y_{12} &= \frac{y_{32}}{m} - \frac{y_{11}}{m} - g\left(y_{11}y_{22}-y_{12}y_{32}\right),
    \label{riccati_y12}
    \\
    \dot y_{21} &= \frac{y_{41}}{m} + m\omega^2y_{22} - g\left(y_{21}^{2}-y_{22}y_{31}\right),
    \label{riccati_y21}
    \\
    \dot y_{22} &= \frac{y_{42}}{m} - \frac{y_{21}}{m} - g\left(y_{21}y_{22}-y_{22}y_{32}\right),
    \label{riccati_y22}
    \\
    \dot y_{31} &= -m\omega^2y_{11} + m\omega^2y_{32} - g\left(y_{31}y_{21}-y_{32}y_{31}\right),
    \label{riccati_y31}
    \\
    \dot y_{32} &= -m\omega^2y_{12} - \frac{y_{31}}{m} - g\left(y_{31}y_{22}-y_{32}^{2}\right),
    \label{riccati_y32}
    \\
    \dot y_{41} &= -m\omega^2y_{21} + m\omega^2y_{42} - g\left(y_{41}y_{21}-y_{42}y_{31}\right),
    \label{riccati_y41}
    \\
    \dot y_{42} &= -m\omega^2y_{22} - \frac{y_{41}}{m} - g\left(y_{41}y_{22}-y_{42}y_{32}\right) - g.
    \label{riccati_y42}
\end{align}

The initial conditions follow from
\begin{equation}
    S(0)=I_6,
    \qquad
    A(0)=I_2,
    \qquad
    C(0)=0,
\end{equation}
so that
\begin{equation}
    Y(0)=0,
\end{equation}
or equivalently,
\begin{equation}
    y_{ij}(0)=0,
    \qquad
    i=1,\ldots,4,
    \qquad
    j=1,2.
    \label{riccati_initial_conditions}
\end{equation}

Thus, the harmonic oscillator problem is reduced to eight autonomous coupled differential equations as Eqs.~\eqref{riccati_y11}--\eqref{riccati_y42}. As shown in the next appendix, the physical trajectory lies on an invariant submanifold of this exact Riccati flow, which allows us to reduce the system further and obtain a fully analytical solution.

\section{Invariant manifold and reduction of the Riccati flow}
\label{riccati_reduction}

In this appendix, we show that the physical solution of the full eight-dimensional system derived in Appendix~\ref{full_riccati} lies on a four-dimensional invariant submanifold. We use this information to reduce the Riccati system further, ultimately arriving at the two complex equations derived in Sec.~\ref{harmonic_riccati_formulation}.

We begin by defining the four quantities
\begin{equation}
\begin{gathered}
    \Phi_1=y_{11}+y_{42},
    \qquad
    \Phi_2=y_{21}+y_{32},\\
    \Phi_3=m^2\omega^2y_{12}-y_{41},
    \qquad
    \Phi_4=m^2\omega^2y_{22}-y_{31}.
\end{gathered}
\end{equation}
Collecting them into
\begin{equation}
    \Phi(Y) =
    \begin{pmatrix}
        \Phi_1\\
        \Phi_2\\
        \Phi_3\\
        \Phi_4
    \end{pmatrix},
\end{equation}
we define the submanifold
\begin{equation}
    \mc{M}=\left\{Y:\Phi(Y)=0\right\}.
    \label{invariant_manifold_definition}
\end{equation}
We note that, on $\mc{M}$, the eight components of $Y$ are completely determined by the four variables $y_{11},y_{12},y_{21}$, and $y_{22}$.

To establish that $\mc{M}$ is invariant under the Riccati flow, we calculate
\begin{equation}
\begin{gathered}
    \dot\Phi_1 = \dot y_{11}+\dot y_{42},
    \qquad
    \dot\Phi_2 = \dot y_{21}+\dot y_{32},\\
    \dot\Phi_3 = m^2\omega^2\dot y_{12}-\dot y_{41},
    \qquad
    \dot\Phi_4 = m^2\omega^2\dot y_{22}-\dot y_{31}.
    \label{invariant_relations_2}
\end{gathered}
\end{equation}

Using Eqs.~\eqref{riccati_y11} and \eqref{riccati_y42}, and imposing Eq.~\eqref{invariant_relations_2}, we find
\begin{equation}
\begin{split}
    \left.\dot\Phi_1\right|_{\mc M}&=m\omega^2(y_{12}+y_{22})-g\left(y_{11}y_{21}-m^2\omega^2y_{12}y_{22}\right)+g\\
    &\quad
    -m\omega^2(y_{12}+y_{22})-g\left(m^2\omega^2y_{12}y_{22}-y_{11}y_{21}\right)-g\\
    &=0.
    \label{phi1_invariance}
\end{split}
\end{equation}
Similarly, from Eqs.~\eqref{riccati_y21} and \eqref{riccati_y32},
\begin{equation}
\begin{split}
    \left.\dot\Phi_2\right|_{\mc M}&=m\omega^2(y_{12}+y_{22})-g\left(y_{21}^2-m^2\omega^2y_{22}^2\right)\\
    &\quad
    -m\omega^2(y_{12}+y_{22})-g\left(m^2\omega^2y_{22}^2-y_{21}^2\right)\\
    &=0.
    \label{phi2_invariance}
\end{split}
\end{equation}
For the third component, Eqs.~\eqref{riccati_y12} and \eqref{riccati_y41} give
\begin{equation}
\begin{split}
    \left.\dot\Phi_3\right|_{\mc M}=&-m\omega^2(y_{11}+y_{21})\\
    &-gm^2\omega^2\left(y_{11}y_{22}+y_{12}y_{21}\right)\\
    &+m\omega^2(y_{11}+y_{21})\\
    &
    +gm^2\omega^2\left(y_{11}y_{22}+y_{12}y_{21}\right) \\
    =&\,\,0.
    \label{phi3_invariance}
\end{split}
\end{equation}
Finally,
\begin{equation}
\begin{split}
    \left.\dot\Phi_4\right|_{\mc M} =& -m\omega^2(y_{11}+y_{21}) -2gm^2\omega^2y_{21}y_{22}\\
    &+m\omega^2(y_{11}+y_{21}) +2gm^2\omega^2y_{21}y_{22}\\
    =& \,\,0.
    \label{phi4_invariance}
\end{split}
\end{equation}
Combining Eqs.~\eqref{phi1_invariance}--\eqref{phi4_invariance}, we obtain
\begin{equation}
    \left.\dot\Phi\right|_{\mc M}=0.
\end{equation}
The physical Riccati vector field is, therefore, tangent to $\mc M$. Hence, $\mc M$ is an invariant submanifold of the full eight-dimensional flow.

The physical initial condition derived in Appendix~\ref{full_riccati} is $Y(0)=0$. Consequently, $\Phi(Y(0))=0$. Thus, the physical trajectory starts on $\mc M$, and hence, it remains on $\mc M$ throughout the interval for which the solution exists. Therefore, along the physical trajectory,
\begin{equation}
\begin{gathered}
    y_{42}(t)=-y_{11}(t),
    \qquad
    y_{32}(t)=-y_{21}(t),\\
    y_{41}(t)=m^2\omega^2y_{12}(t),
    \qquad
    y_{31}(t)=m^2\omega^2y_{22}(t).
    \label{physical_invariant_relations}
\end{gathered}
\end{equation}

We can now substitute Eq.~\eqref{physical_invariant_relations} into the first four equations of the full Riccati system. Eqs.~\eqref{riccati_y11}--\eqref{riccati_y22} become
\begin{gather}
    \dot y_{11} = m\omega^2(y_{12}+y_{22}) -gy_{11}y_{21} +gm^2\omega^2y_{12}y_{22} +g,
    \label{reduced_y11}\\
    \dot y_{12} = -\frac{y_{11}+y_{21}}{m} -g\left( y_{11}y_{22} +y_{12}y_{21} \right).
    \label{reduced_y12}\\
    \dot y_{21} = m\omega^2(y_{12}+y_{22}) -gy_{21}^2 +gm^2\omega^2y_{22}^2,
    \label{reduced_y21}\\
    \dot y_{22} = -\frac{y_{11}+y_{21}}{m} -2gy_{21}y_{22}.
    \label{reduced_y22}
\end{gather}
Thus, the physical Riccati dynamics is governed by the closed four-dimensional system Eqs.~\eqref{reduced_y11}--\eqref{reduced_y22}, with
\begin{equation}
    y_{11}(0) = y_{12}(0) = y_{21}(0) = y_{22}(0) = 0.
    \label{reduced_initial_conditions}
\end{equation}

Having reduced the Riccati equations to a four-dimensional system, we now introduce
\begin{equation}
    u=y_{11}-y_{21},
    \qquad
    v=y_{12}-y_{22}.
    \label{uv_definition}
\end{equation}
Subtracting Eq.~\eqref{reduced_y21} from Eq.~\eqref{reduced_y11}, we obtain
\begin{equation}
    \dot u=g\left[1-y_{21}u+m^2\omega^2y_{22}v\right].
    \label{u_equation}
\end{equation}
Similarly, subtracting Eq.~\eqref{reduced_y22} from Eq.~\eqref{reduced_y12} gives
\begin{equation}
    \dot v=-g\left(y_{22}u+y_{21}v\right).
    \label{v_equation}
\end{equation}

We now combine the variables into the complex quantities
\begin{equation}
    z = u+im\omega v,
    \qquad
    Z = y_{21}+im\omega y_{22}.
    \label{zZ_definition_appendix}
\end{equation}
Their product is
\begin{equation}
\begin{split}
    Zz&=\left(y_{21}+im\omega y_{22}\right)\left(u+im\omega v\right)\\
    &=y_{21}u-m^2\omega^2y_{22}v+im\omega\left(y_{21}v+y_{22}u\right).
    \label{Zz_product}
\end{split}
\end{equation}
Using Eqs.~\eqref{u_equation}, \eqref{v_equation} and \eqref{Zz_product}, we therefore obtain
\begin{equation}
    \dot z = \dot u+im\omega\dot v = g(1-Zz).
    \label{z_complex_equation_appendix}
\end{equation}

For the second complex variable,
\begin{equation}
    \dot Z = \dot y_{21} + im\omega\dot y_{22}.
\end{equation}
Using Eqs.~\eqref{reduced_y21} and \eqref{reduced_y22},
\begin{equation}
\begin{split}
    \dot Z &= m\omega^2(y_{12}+y_{22}) -i\omega(y_{11}+y_{21})\\
    &\quad
    -g\left(y_{21}^2-m^2\omega^2y_{22}^2+2im\omega y_{21}y_{22}\right).
\end{split}
\end{equation}
Since
\begin{equation}
    Z^2=y_{21}^2-m^2\omega^2y_{22}^2+2im\omega y_{21}y_{22},
\end{equation}
and, from Eq.~\eqref{uv_definition},
\begin{equation}
    -2i\omega Z-i\omega z=m\omega^2(y_{12}+y_{22})-i\omega(y_{11}+y_{21}),
\end{equation}
we finally obtain
\begin{equation}
    \dot Z = -gZ^2 -2i\omega Z -i\omega z.
    \label{Z_complex_equation_appendix}
\end{equation}

So, the eight-dimensional Riccati dynamics restricted to the physical invariant manifold is equivalent to
\begin{equation}
    \dot z=g(1-Zz),
    \qquad
    \dot Z=-gZ^2-2i\omega Z-i\omega z,
    \label{two_complex_system_appendix}
\end{equation}
with
\begin{equation}
    z(0)=Z(0)=0.
\end{equation}

In the next appendix, the coupled nonlinear system described by Eq.~\eqref{two_complex_system_appendix} is reduced further to a single linear third-order differential equation.

\section{Linearization and spectral structure}
\label{linearization_spectral_structure}

In this appendix, we reduce the two complex variable system derived in
Appendix~\ref{riccati_reduction} to a linear differential equation of order three, and analyze its spectrum. We begin from
\begin{equation}
    \dot z=g(1-Zz),
    \qquad
    \dot Z=-gZ^2-2i\omega Z-i\omega z,
    \label{complex_system_D}
\end{equation}
with the physical initial conditions
\begin{equation}
    z(0)=Z(0)=0.
    \label{complex_initial_D}
\end{equation}

We first eliminate $z$. From Eq.~\eqref{complex_system_D}, 
\begin{equation}
    z = \frac{i}{\omega}\dot Z + \frac{ig}{\omega}Z^2 - 2Z.
    \label{z_in_terms_of_Z}
\end{equation}
Differentiating, 
\begin{equation}
    \dot z = \frac{i}{\omega}\ddot Z + \frac{2ig}{\omega}Z\dot Z - 2\dot Z.
    \label{zdot_in_terms_of_Z}
\end{equation}
On the other hand, substituting Eq.~\eqref{z_in_terms_of_Z} into the first
equation in Eq.~\eqref{complex_system_D}
\begin{equation}
    \dot z = g - \frac{ig}{\omega}Z\dot Z - \frac{ig^2}{\omega}Z^3 + 2gZ^2.
    \label{zdot_second_form}
\end{equation}
Equating Eqs.~\eqref{zdot_in_terms_of_Z} and \eqref{zdot_second_form}, we obtain
\begin{equation}
    \ddot Z + 3gZ\dot Z + 2i\omega\dot Z + g^2Z^3 + 2ig\omega Z^2 + ig\omega = 0.
    \label{nonlinear_Z_equation}
\end{equation}
Thus, the original coupled first-order equation is equivalent to a single second-order nonlinear equation for $Z(t)$.

Eq.~\eqref{nonlinear_Z_equation} can be further reduced to a linear equation through the substitution
\begin{equation}
    Z(t) = \frac{\dot{\mc W}(t)} {g\mc W(t)}.
    \label{Z_W_transformation}
\end{equation}
For $g\neq0$ and on an interval where $\mc W(t)\neq0$ (the next appendix shows that this is true for all finite times), this substitution gives
\begin{equation}
\begin{gathered}
    \dot Z+gZ^2=\frac{\ddot{\mc W}}{g\mc W},\\
    \ddot Z+3gZ\dot Z+g^2Z^3 =\frac{\dddot{\mc W}}{g\mc W}.
\end{gathered}
\end{equation}
Thus, the above substitution reduces the Eq.~\eqref{nonlinear_Z_equation} to
\begin{equation}
    \dddot{\mc W} + 2i\omega\ddot{\mc W} + ig^2\omega\mc W = 0.
    \label{third_order_W}
\end{equation}
Now, we have succeeded in encoding the full physical Riccati flow into a single linear third-order differential equation.

Since $Z \propto \dot{\mc W}/{\mc W}$, the overall normalization of $\mc W$ is arbitrary. Hence, we choose $\mc W(0)=1$. From $Z(0)=0$ and Eq.~\eqref{Z_W_transformation}, $\dot{\mc W}(0)=0$. Furthermore, the second equation in Eq.~\eqref{complex_system_D} gives $\dot Z(0)=0$, because $Z(0)=z(0)=0$. Differentiating
Eq.~\eqref{Z_W_transformation} then gives $\ddot{\mc W}(0)=0$. Thus, the physical initial conditions for the linear problem are
\begin{equation}
    \mc W(0)=1,
    \qquad
    \dot{\mc W}(0)=0,
    \qquad
    \ddot{\mc W}(0)=0.
    \label{W_initial_conditions_appendix}
\end{equation}

We now turn to the spectrum of Eq.~\eqref{third_order_W}. Taking $\mc W(t)=e^{rt}$, we obtain the characteristic equation as
\begin{equation}
    r^3+2i\omega r^2+ig^2\omega=0.
    \label{r_characteristic_D}
\end{equation}
Introducing $r=-i\lambda$, the characteristic equation becomes
\begin{equation}
    P(\lambda) = \lambda^3 - 2\omega\lambda^2 + g^2\omega = 0,
    \label{lambda_cubic_D}
\end{equation}
with discriminant
\begin{equation}
    \Delta=g^2\omega^2\left(32\omega^2-27g^2\right).
    \label{discriminant_D}
\end{equation}
Thus, the nature of the characteristic roots changes at
\begin{equation}
    g_c = \sqrt{\frac{32}{27}}\, \omega,
    \label{gc_D}
\end{equation}
where we take $g>0$ without loss of generality. 

For $0<g<g_c$, the discriminant $\Delta > 0$. Hence, the cubic Eq.~\eqref{lambda_cubic_D} has three distinct real roots,
\begin{equation}
    \lambda_1,\lambda_2,\lambda_3\in\mathbb{R}.
\end{equation}
Consequently, 
\begin{equation}
    r_j=-i\lambda_j,
    \qquad
    j=1,2,3,
\end{equation}
are all purely imaginary. Thus, no exponentially growing or decaying characteristic mode is present in this regime.

Secondly, at the critical coupling, 
\begin{equation} 
g=g_c, 
\end{equation} 
the cubic factorizes as 
\begin{equation} 
    P(\lambda) = \left( \lambda+\frac{2\omega}{3} \right) \left( \lambda-\frac{4\omega}{3} \right)^2. 
\label{critical_factorization} 
\end{equation} 
Thus, \begin{equation} 
    \lambda_0=-\frac{2\omega}{3}, 
    \qquad 
    \lambda_c=\frac{4\omega}{3}, 
\end{equation} 
where $\lambda_c$ is a repeated root. The associated characteristic exponents are 
\begin{equation} 
    r_0=\frac{2i\omega}{3}, 
    \qquad 
    r_c=-\frac{4i\omega}{3}. 
\end{equation} 
The solution satisfying the physical initial conditions Eq.~\eqref{W_initial_conditions_appendix} is 
\begin{equation} 
    \mc W(t) = \f{4}{9}e^{2i\omega t/3} + \left( \f{5}{9}+\f{4i\omega}{9}t \right)e^{-4i\omega t/3}. 
\label{critical_general_solution} 
\end{equation} 
Thus, although the characteristic exponents still have vanishing real parts, the repeated root produces a polynomial prefactor, and the critical case is distinct from the generic subcritical regime. 

Lastly, for $g>g_c$, the discriminant is negative, and the cubic has one real root and one complex-conjugate pair:
\begin{equation}
    \lambda_0\in\mathbb{R},
    \quad
    \lambda_+=\alpha+i\beta,
    \quad
    \lambda_-=\alpha-i\beta,
    \quad
    \beta>0.
    \label{lambda_complex_pair_D}
\end{equation}
The corresponding characteristic exponents are
\begin{equation}
    r_0=-i\lambda_0,
    \qquad
    r_+=\beta-i\alpha,
    \qquad
    r_-=-\beta-i\alpha.
    \label{r_complex_pair_D}
\end{equation}

Thus, in the regime $g>g_c$, the appearance of the expanding--contracting pair is the origin of the hyperbolic mechanism. However, the existence of an eigenvalue with a positive real part is not sufficient for determining the long-term behavior of the solution, because its coefficient could, in principle, vanish. In the next appendix, therefore, we show that the growing mode has a non-vanishing coefficient, and hence, dominates the long-time behavior of the system.

\section{Nonvanishing of the auxiliary function $\mc W(t)$}
\label{nonvanishing_W}

The transformation
\begin{equation}
    Z(t)=\frac{\dot{\mc W}(t)}{g\mc W(t)}
    \label{W_transformation_nonzero}
\end{equation}
used in Appendix~\ref{linearization_spectral_structure} is well defined
only if $\mc W(t)\neq0$. In this appendix, we show that the solution
selected by the physical initial conditions does not vanish at any finite time.

The characteristic equation derived in Appendix~\ref{linearization_spectral_structure} is
\begin{equation}
    r^3+2i\omega r^2+ig^2\omega=0.
\end{equation}
We introduced $r=-i\lambda$. For the present proof, it is convenient to define $\nu=-\lambda$, that is $r=i\nu$. Thus, the cubic equation takes the form
\begin{equation}
    p(\nu) = \nu^3+2\omega\nu^2-g^2\omega = 0.
    \label{nu_cubic_nonzero}
\end{equation}
For $g\neq0$ and $\omega>0$,
\begin{equation}
    p(0)=-g^2\omega<0,
    \qquad
    p(\nu)\longrightarrow+\infty
    \quad
    \text{as } \nu\longrightarrow+\infty.
\end{equation}
Moreover,
\begin{equation}
    p'(\nu) = \nu(3\nu+4\omega)>0
    \qquad
    \text{for }\nu>0.
\end{equation}
Hence, the cubic possesses exactly one positive real root, for all $g\ne 0$, $\omega> 0$, which we denote by
\begin{equation}
    \nu_1>0.
    \label{positive_root_nonzero}
\end{equation}

When the three roots are distinct, the general solution can be written
as
\begin{equation}
    \mc W(t)=\sum_{j=1}^{3}c_j e^{i\nu_jt}.
    \label{W_general_nonzero}
\end{equation}
Solving for the coefficients gives
\begin{equation}
    c_j=\frac{\nu_k\nu_l}{(\nu_j-\nu_k)(\nu_j-\nu_l)},
    \qquad
    \{j,k,l\}=\{1,2,3\}.
    \label{coefficients_nonzero}
\end{equation}
Using the characteristic equation, the $c_1$ corresponding to the positive root can be found as
\begin{equation}
    c_1=\frac{\nu_1+2\omega}{3\nu_1+4\omega}.
    \label{c1_nonzero}
\end{equation}
From the above expression, we find
\begin{equation}
    \frac{1}{2}-c_1=\frac{\nu_1}{2(3\nu_1+4\omega)}.
\end{equation}
Since $\nu_1 > 0$, $\frac{1}{2}-c_1 > 0$. Hence, 
\begin{equation}
    0<c_1<\frac{1}{2},
    \label{c1_bound_nonzero}
\end{equation}

Now, we consider the three possible regimes. First, consider $g<g_c$. Since $p\!\left(-\frac{4\omega}{3}\right)>0$, the ordering of the roots are
\begin{equation}
    \nu_3 < -\frac{4\omega}{3} < \nu_2 < 0 < \nu_1.
    \label{real_root_order_nonzero}
\end{equation}
For each real root,
\begin{equation}
    c_j=\frac{\nu_j+2\omega}{3\nu_j+4\omega}.
    \label{real_coefficients_nonzero}
\end{equation}
It follows that
\begin{equation}
    c_1>0,
    \qquad
    c_2>0,
    \qquad
    c_3<0.
\end{equation}
Since $c_1+c_2+c_3=1$,
\begin{equation}
    c_2-|c_3|=c_2+c_3=1-c_1.
\end{equation}
Using the reverse triangle inequality on
Eq.~\eqref{W_general_nonzero},
\begin{equation}
\begin{split}
    |\mc W(t)| &\ge \left| c_2e^{i\nu_2t} + c_3e^{i\nu_3t} \right| - c_1\\
    &\ge c_2-|c_3|-c_1\\
    &= 1-2c_1\\
    &= \frac{\nu_1}{3\nu_1+4\omega} >0.
    \label{subcritical_bound_nonzero}
\end{split}
\end{equation}

Now, for $g^2>g_c^2$, there is one real root $\nu_1>0$ and a complex-conjugate pair,
\begin{equation}
    \nu_2=\alpha+i\beta,
    \qquad
    \nu_3=\alpha-i\beta,
    \qquad
    \beta>0.
\end{equation}
Since the differential equation and the initial conditions imply the
corresponding conjugation relation,
\begin{equation}
    c_3=\overline{c_2},
\end{equation}
the solution takes the form
\begin{equation}
    \mc W(t)=c_1e^{i\nu_1t}+e^{i\alpha t}\left(c_2e^{-\beta t}+\overline{c_2}e^{\beta t}\right).
    \label{hyperbolic_W_nonzero}
\end{equation}
Writing $c_2=a+ib$, the magnitude of the term in parentheses satisfies
\begin{equation}
\begin{split}
    \left|c_2e^{-\beta t}+\overline{c_2}e^{\beta t}\right|^2&=4a^2\cosh^2(\beta t)+4b^2\sinh^2(\beta t)\\
    &\ge4a^2.
\end{split}
\end{equation}
Hence,
\begin{equation}
    \left|c_2e^{-\beta t}+\overline{c_2}e^{\beta t}\right|\ge2|\on{Re}c_2|.
\end{equation}
Since $c_1+c_2+\overline{c_2}=1$, we have $2\on{Re}c_2=1-c_1>0$. Therefore,
\begin{equation}
\begin{split}
    |\mc W(t)| &\ge 2|\on{Re}c_2|-c_1\\
    &= 1-2c_1\\
    &= \frac{\nu_1}{3\nu_1+4\omega} >0.
    \label{hyperbolic_bound_nonzero}
\end{split}
\end{equation}

Finally, at the critical value $g^2=\frac{32}{27}\omega^2$, the roots are
\begin{equation}
    \nu_1=\frac{2\omega}{3},
    \qquad
    \nu_2=\nu_3=-\frac{4\omega}{3}.
\end{equation}
As found in Appendix~\ref{linearization_spectral_structure}, the physical
solution is
\begin{equation}
    \mc W(t) = \frac{4}{9}e^{2i\omega t/3} + \left( \frac{5}{9} + \frac{4i\omega}{9}t \right) e^{-4i\omega t/3}.
    \label{critical_W_nonzero}
\end{equation}
Using the reverse triangle inequality,
\begin{equation}
\begin{split}
    |\mc W(t)| &\ge \left| \frac{5}{9} + \frac{4i\omega}{9}t \right| - \frac{4}{9}\\
    &= \sqrt{ \left(\frac{5}{9}\right)^2 + \left(\frac{4\omega t}{9}\right)^2 } - \frac{4}{9}\\
    &\ge \frac{1}{9} >0.
    \label{critical_bound_nonzero}
\end{split}
\end{equation}

Combining Eqs.~\eqref{subcritical_bound_nonzero},
\eqref{hyperbolic_bound_nonzero}, and
\eqref{critical_bound_nonzero}, we obtain, for $g\neq0$ and
$\omega>0$,
\begin{equation}
    |\mc W(t)| \ge \frac{\nu_1}{3\nu_1+4\omega} >0
    \qquad
    \text{for all finite }t.
    \label{universal_nonzero_bound}
\end{equation}

At $g=0$, Eq.~\eqref{third_order_W}, along with the physical initial conditions, gives $\mc W(t)=1$. So, the auxiliary function is nonzero in that case as well. 

To conclude, $\mc W(t)$ has no zeros for finite time. In particular, for
$g\neq0$, the
\begin{equation}
    Z(t) = \frac{\dot{\mc W}(t)}
    {g\mc W(t)}
\end{equation}
does not develop a singularity through a zero of $\mc W(t)$. Hence, the above substitution used to solve the physical Riccati solution is well-defined for all finite times.

\section{Physical mode selection and asymptotic fixed point}
\label{physical_mode_selection}

In Appendix~\ref{linearization_spectral_structure}, we showed that, for $g>g_c$, the spectrum contains one purely imaginary characteristic exponent together with an expanding--contracting pair, as given in Eq.~\eqref{r_complex_pair_D}. The exponent $r_+$ has the positive real part. In this appendix, we show that the coefficient of the growing mode is nonzero for the physical initial conditions, thereby deducing the long-time behavior.

The general solution of Eq.~\eqref{third_order_W} in the regime $g>g_c$ is
\begin{equation}
    \mc W(t) = c_0e^{-i\lambda_0t} + c_+e^{-i\lambda_+t} +  c_-e^{-i\lambda_-t}.
    \label{general_W_E}
\end{equation}
Under the physical initial conditions $\mc W(0)=1, \dot{\mc W}(0)=0, \ddot{\mc W}(0)=0$,
\begin{gather}
    c_0+c_++c_-=1,
    \label{coefficient_condition_1}\\
    \lambda_0c_0+\lambda_+c_++\lambda_-c_-=0,
    \label{coefficient_condition_2}\\
    \lambda_0^2c_0+\lambda_+^2c_+ +\lambda_-^2c_-=0.
    \label{coefficient_condition_3}
\end{gather}
Solving these equations for the coefficient of the growing mode gives
\begin{equation}
    c_+ = \frac{\lambda_0\lambda_-} {(\lambda_+-\lambda_0)(\lambda_+-\lambda_-)}.
    \label{growing_coefficient}
\end{equation}

The characteristic polynomial is
\begin{equation}
    P(\lambda) = \lambda^3-2\omega\lambda^2+g^2\omega.
\end{equation}
Since the constant term of the characteristic polynomial is nonzero, none of its roots can vanish. Also, for $g>g_c$, the three roots are distinct. So, $\lambda_+-\lambda_0\neq0$ and $\lambda_+-\lambda_-\neq0$. Therefore
\begin{equation}
    c_+\neq0.
    \label{nonzero_growing_coefficient}
\end{equation}
Thus, the initial conditions imply that the growing mode is present in the solution.

Using Eq.~\eqref{general_W_E}, the general solution can be written as
\begin{equation}
    \mc W(t) = c_+e^{(\beta-i\alpha)t} + c_0e^{-i\lambda_0t} + c_-e^{(-\beta-i\alpha)t}.
    \label{W_modes_E}
\end{equation}
Since $\beta>0$ and $c_+\neq0$, the first term dominates at long times. Factoring it out gives
\begin{equation}
    \mc W(t) = c_+e^{(\beta-i\alpha)t} \left[ 1+O(e^{-\beta t}) \right],
    \qquad
    t\longrightarrow\infty.
    \label{W_asymptotic_E}
\end{equation}
Consequently,
\begin{equation}
    Z(t) = \frac{\dot{\mc W}(t)}{g\mc W(t)} = \f{\beta-i\alpha}{g} + O(e^{-\beta t}).
    \label{logarithmic_asymptotic_E}
\end{equation}
Hence, $Z(t)$ approaches the asymptotic fixed point
\begin{equation}
    Z_* = \frac{\beta-i\alpha}{g}.
    \label{Z_fixed_point_E}
\end{equation}
From the definition $Z = y_{21} + im\omega y_{22}$, the real and imaginary parts of Eq.~\eqref{Z_fixed_point_E} give
\begin{equation}
    y_{21}(t) \longrightarrow y_{21}^* = \frac{\beta}{g},
\quad 
y_{22}(t) \longrightarrow y_{22}^* = -\frac{\alpha}{gm\omega}.
    \label{y_fixed_E}
\end{equation}
In particular,
\begin{equation}
    gy_{21}^*=\beta>0.
    \label{positive_y21_E}
\end{equation}

Thus, the physical initial conditions select the expanding characteristic mode given by the exponent $r_+$ in the hyperbolic regime. The corresponding Riccati trajectory approaches a fixed point with strictly positive $gy_{21}$. This provides the rate at which $A^{-1}(t)$ decays in the long-time limit, as shown in the next appendix.

\section{Exact spectral solution for $A^{-1}(t)$}
\label{spectral_Ainverse}

In this appendix, we use the governing equation of $A^{-1}(t)$, and the asymptotic fixed point found in Appendix~\ref{physical_mode_selection} to determine its long-time behavior. From the Riccati formulation,
\begin{equation}
    \dot A^{-1} = -A^{-1}K(t), \qquad K(t) = M_{11}+M_{12}Y(t).
    \label{Ainverse_equation_F}
\end{equation}
Using the matrices introduced in Appendix~\ref{full_riccati}, and the invariant submanifold derived in Appendix~\ref{riccati_reduction},
\begin{equation}
    K(t) =
    \begin{pmatrix}
        gy_{21}
        &
        \dfrac{1}{m}+gy_{22}
        \\[2mm]
        -m\omega^2-gm^2\omega^2y_{22}
        &
        gy_{21}
    \end{pmatrix}.
    \label{K_invariant_F}
\end{equation}

Equation~\eqref{K_invariant_F} can be written as
\begin{equation}
    K(t) = a(t)I_2+b(t)J,
    \label{K_IJ_F}
\end{equation}
where $a(t)=gy_{21}(t)$, $b(t)=\frac{1}{m}+gy_{22}(t)$, $I_2$ is the identity matrix of dimension $2\times 2$, and $J = \begin{pmatrix} 0 & 1\\ -m^2\omega^2 & 0 \end{pmatrix}.$ Because $K(t)$ is a linear combination of $I_2$ and a time-independent matrix $J$, the eigenvectors of $K(t)$ remain the same for all times. It follows immediately that 
\begin{equation}
    [K(t_1),K(t_2)]=0
    \qquad
    \forall\,t_1,t_2.
    \label{K_commutativity_F}
\end{equation}
Hence, the solution of $A^{-1}(t)$ will not involve a time-ordering.

The two eigenvalues of $K(t)$ are
\begin{equation}
    \kappa_\pm(t) = gy_{21}(t) \pm i\omega \left[ 1+gmy_{22}(t) \right].
    \label{kappa_F}
\end{equation}

The corresponding spectral projectors are
\begin{equation}
    P_\pm = \frac{1}{2}
    \begin{pmatrix}
        1 & \mp\dfrac{i}{m\omega}\\[2mm]
        \pm im\omega & 1
    \end{pmatrix}.
    \label{explicit_projectors_F}
\end{equation}
Thus,
\begin{equation}
    K(t) = \kappa_+(t)P_+ + \kappa_-(t)P_-.
    \label{K_spectral_F}
\end{equation}

 Since $A^{-1}(0)=I_2$, the solution of $A^{-1}(t)$ is
\begin{equation}
    A^{-1}(t) = \exp \left[ -\int_0^t K(s)\,ds \right].
    \label{Ainverse_exponential_F}
\end{equation}
Using Eq.~\eqref{K_spectral_F} and the time-independent projectors,
\begin{equation}
    A^{-1}(t) = \sum_{\sigma=\pm} \exp \left[ -\int_0^t \kappa_\sigma(s)\,ds \right]  P_\sigma.
    \label{Ainverse_spectral_F}
\end{equation}
Defining
\begin{equation}
    \Theta(t) = \omega \int_0^t \left[ 1+gmy_{22}(s) \right]ds,
    \label{Theta_F}
\end{equation}
we may write
\begin{equation}
    A^{-1}(t) = \exp \left[ -g\int_0^t y_{21}(s)\,ds \right] \left[ e^{-i\Theta(t)}P_+ + e^{i\Theta(t)}P_- \right].
    \label{Ainverse_phase_F}
\end{equation}
The phase $\Theta(t)$ produces an oscillatory contribution, but the envelope of $A^{-1}(t)$ is controlled by the factor
\begin{equation}
    \exp \left[ -g\int_0^t y_{21}(s)\,ds \right].
    \label{Ainverse_envelope_F}
\end{equation}

In Appendix~\ref{physical_mode_selection}, we showed that in the hyperbolic regime $g>g_c$, the $y_{21}$ approaches a positive constant $\beta/g$. Since 
\begin{equation} 
y_{21}(t) = \frac{\beta}{g} + O(e^{-\beta t}), 
\end{equation} 
we have 
\begin{equation} 
g\int_0^t y_{21}(s)\,ds = \beta t+O(1). 
\end{equation} 
Therefore,
\begin{equation}
    \exp \left[ -g\int_0^t y_{21}(s)\,ds  \right] = O(e^{-\beta t}).
    \label{envelope_decay_F}
\end{equation}
Since the remaining factors in Eq.~\eqref{Ainverse_phase_F} are oscillatory, and the projectors are time-independent, we obtain
\begin{equation}
    A^{-1}(t)\longrightarrow0,
    \qquad
    t\longrightarrow\infty.
    \label{Ainverse_zero_F}
\end{equation}

Thus, in the hyperbolic regime $g>g_c$, $A^{-1}(t)$ decays because both its spectral components have the same decaying envelope. The decay rate is fixed by the positive quantity $gy_{21}^*=\beta$, which was dynamically selected by the Riccati system. This establishes the contraction of every nonzero subsystem separation as discussed in the Sec.~\ref{harmonic_Wigner_flattening}.

\section{Spectrum of the full Hamiltonian generator}
\label{spectrum_of_M}

Here, we verify that the same critical coupling encountered in the Riccati flow analysis also appears directly in the spectrum of the full $6\times6$ Hamiltonian generator. We denote the eigenvalues of $M$ by $\mu$ in order to distinguish them from the roots $\lambda_j$ introduced in Eq.~\eqref{lambda_cubic_D}.

For the symmetric model, the Hamiltonian generator is given by the Eq.~\eqref{full_harmonic_generator}. The characteristic determinant is given by
\begin{align}
    \det(\mu I_6-M) &= (\mu^2+\omega^2)^3 -2g^2\omega^2(\mu^2+\omega^2) +g^4\omega^2.
    \label{appendix_full_characteristic}
\end{align}
Thus, defining $y=\mu^2+\omega^2$, the sixth-order characteristic equation reduces to the depressed cubic
\begin{equation}
    f(y) = y^3-2g^2\omega^2y+g^4\omega^2 =0.
    \label{appendix_y_cubic}
\end{equation}
The discriminant is
\begin{equation}
\begin{split}
    \Delta_M &= -4(-2g^2\omega^2)^3 -27(g^4\omega^2)^2 \\
    &= g^6\omega^4 \left(32\omega^2-27g^2\right).
    \label{appendix_M_discriminant}
\end{split}
\end{equation}
Hence, for nonzero coupling, the discriminant vanishes at
\begin{equation}
    g_c^2=\frac{32}{27}\omega^2,
    \label{appendix_M_gc}
\end{equation}
which is precisely the critical coupling obtained independently from the Riccati characteristic equation in Eq.~\eqref{lambda_cubic}. The additional zero at $g=0$ corresponds to the non-interacting oscillators case.

For $g<g_c$, Eq.~\eqref{appendix_y_cubic} possesses three real
roots. We note that
\begin{enumerate}
    \item $f(y\to\infty) = \infty$,
    \item $f'(y) =0$, for $y = \pm\sqrt{2/3}\,g\omega$,
    \item $+\sqrt{2/3}\,g\omega < \omega^2$, for all values of $g<g_c$,
    \item $f(+\sqrt{2/3}\,g\omega) < 0$ and $f(-\sqrt{2/3}\,g\omega)> 0$ for $g< g_c$,
    \item $f(+\sqrt{2/3}\,g\omega) > 0$ and $f(-\sqrt{2/3}\,g\omega)> 0$ for $g> g_c$,
    \item $f(\omega^2) = \omega^6 - 2g^2\omega^4 + g^4\omega^2\ge 0$, for all values of $g<g_c$,
    \item $f(\omega^2) = 0$, only when $g = \omega$.
\end{enumerate}
Thus, except when $g=\omega$, the cubic has three distinct real roots $y_i$ ($i=1,2,3$), with $y_i < \omega^2$. The corresponding values of $\mu^2=y_i-\omega^2$ are negative, and the six eigenvalues of $M$ therefore remain purely imaginary. In case of $g=\omega$, one of the roots $\mu$ will be equal to zero, and the other two purely imaginary. The generator thus has a double zero eigenvalue with two independent eigenvectors, together with two nonzero imaginary pairs. The evolution remains bounded, with constant components in the zero-mode subspace and oscillatory components in the others.

At the critical coupling, $g^2=g_c^2$, Eq.~\eqref{appendix_y_cubic} factorizes as
\begin{equation}
    \left( y-\frac{8}{9}\omega^2 \right)^2 \left( y+\frac{16}{9}\omega^2 \right) =0.
    \label{appendix_M_critical_factorization}
\end{equation}
Thus, at the critical point, the spectrum of the full generator is
\begin{equation}
    \mu=\pm\frac{i\omega}{3},
    \qquad
    \mu=\pm\frac{5i\omega}{3},
    \label{appendix_M_critical_eigenvalues}
\end{equation}
where the pair $\mu=\pm i\omega/3$ has algebraic multiplicity two and geometric multiplicity one. The corresponding Jordan blocks produce terms proportional to $t e^{\pm i\omega t/3}$. Thus, the critical dynamics can grow linearly despite having no eigenvalues with nonzero real parts.

For $g>g_c$, the discriminant becomes negative. The cubic in Eq.~\eqref{appendix_y_cubic} then has one negative real root and one complex-conjugate pair. The latter generates four eigenvalues of $M$ with nonzero real parts, appearing as the symplectic quartet
\begin{equation}
    \mu,\quad
    -\mu,\quad
    \mu^*,\quad
    -\mu^*.
\end{equation}
The remaining two eigenvalues form a purely imaginary pair. Thus, the full Hamiltonian generator develops exponential expanding and contracting directions precisely when
\begin{equation}
    g>g_c.
\end{equation}

This calculation clarifies that the critical coupling found through the Riccati reduction has a direct meaning at the level of the complete Hamiltonian flow.

\section{Regularity condition for hyperbolic Wigner flattening}
\label{regularity_wigner_flattening}

In Sec.~\ref{harmonic_Wigner_flattening}, we argued that the decay of $A^{-1}(t)$ causes the Wigner function to become independent of the subsystem coordinates asymptotically, provided that the Wigner function is sufficiently regular. In this appendix, we make the idea of regularity more precise.

From the exact reduced Wigner function, the difference between the reduced Wigner functions at two subsystem points $R_S,R_S'$ is
\begin{equation}
\begin{split}
     &W_S(R_S,t)-W_S(R_S',t)\\
     &\qquad= \frac{1}{|\det A(t)|} \int d^4u\, \Big[ W_0\!\left( A^{-1}R_S-Fu,u \right) \\
    &\qquad\quad - W_0\!\left( A^{-1}R_S'-Fu,u \right) \Big],
    \label{Wigner_difference_G}
\end{split}
\end{equation}
where $F(t)=A^{-1}(t)B(t)$. The two subsystem arguments of the initial Wigner function differ by
\begin{equation}
    \delta q(t) = A^{-1}(t)(R_S-R_S').
    \label{delta_q_G}
\end{equation}
If $R_S$ and $R_S'$ belong to a fixed compact region $K\subset\mathbb{R}^2$, then
\begin{equation}
    \|\delta q(t)\| \le \|A^{-1}(t)\| \on{diam}(K),
    \label{delta_q_bound_G}
\end{equation}
where $\on{diam}(K) \geq \|R_S-R'_S\|$, for $R_S,R'_S\in K$. Hence, in the hyperbolic regime $g>g_c$, $A^{-1}(t)\to 0$ implies that the separation between the two subsystem arguments in Eq.~\eqref{Wigner_difference_G} tends to zero for all $R_S,R'_S\in K$.

A sufficient regularity condition is obtained by defining
\begin{equation}
    \Omega_0(\delta) = \int d^4u\, \sup_{\substack{q\in\mathbb{R}^2\\
    \|h\|\le\delta}} \left| W_0(q+h,u)-W_0(q,u) \right|.
    \label{modulus_G}
\end{equation}
Here $q$ and $h$ are two-component subsystem phase-space vectors. We assume that $\Omega_0(\delta)<\infty$ for sufficiently small $\delta$, and that
\begin{equation}
    \Omega_0(\delta) \longrightarrow0
    \qquad
    \text{as } \delta\longrightarrow0.
    \label{regularity_condition_G}
\end{equation}
This condition controls the change in $W_0$ under small subsystem displacements, uniformly over the subsystem coordinates, after integrating out the environment coordinates. It therefore gives a bound on the variation of the reduced Wigner function. 

Using Eq.~\eqref{DeltaK} and Eqs.~\eqref{Wigner_difference_G}--\eqref{delta_q_bound_G}, we obtain
\begin{equation}
\begin{split}
    &\sup_{R_S,R_S'\in K} \left| W_S(R_S,t)-W_S(R_S',t) \right| \\
    &\qquad\le \frac{1}{|\det A(t)|} \Omega_0\!\left( \|A^{-1}(t)\| \on{diam}(K) \right).
    \label{flattening_bound_G}
\end{split}
\end{equation}
Since 
\begin{equation} 
\frac{1}{|\det A(t)|} = |\det A^{-1}(t)| \longrightarrow 0
\end{equation}
and, by Eq.~\eqref{regularity_condition_G}, 
\begin{equation}
    \Omega_0\!\left( \|A^{-1}(t)\| \on{diam}(K) \right) \longrightarrow0,
\end{equation}
we can conclude that
\begin{equation}
    \sup_{R_S,R_S'\in K} \left| W_S(R_S,t)-W_S(R_S',t) \right| \longrightarrow0.
    \label{local_flattening_G}
\end{equation}
Thus, under the regularity condition Eq.~\eqref{regularity_condition_G}, the decay of $A^{-1}(t)$ is sufficient to establish local Wigner flattening on every fixed compact subsystem region.

\section{Confinement to a finite region}
\label{finite_volume_robustness}

At sufficiently strong coupling, the original quadratic Hamiltonian becomes unbounded from below. In this section, we show that confining all three position coordinates to a sufficiently large box restores a lower bound on the Hamiltonian while preserving the continuum dynamics to arbitrary, controlled precision over any fixed time interval. We restrict each coordinate to the interval $(-L, L)$ and impose a Dirichlet boundary condition where the wavefunctions vanish at the boundary. A Hamiltonian $H_L$, with the same form as $H_{\mathrm{ho}}$, is defined to act on the confined space with the Dirichlet boundary condition. 

By completing the momentum squares, the Hamiltonian takes the form
\begin{equation}
\begin{split}
    H_L = & \frac{(p + mgx_2)^2 + (p_1 + mgx)^2 + p_2^2}{2m}\\
    &+ \frac{m\omega^2 x_1^2}{2} + \frac{m}{2}(\omega^2 - g^2)(x^2 + x_2^2).
\end{split}
\end{equation}
The squared terms have nonnegative expectation values. The last term can be negative, but $x^2+x_2^2\leq2L^2$. Therefore, $H_L$ is thus bounded below for every finite $L$.

We next have to see how the walls change the evolution. Let $\psi_0\in\mc S(\mathbb R^3)$ be normalized. The continuum evolution is by the $H_{\mr{ho}}$ as
\begin{equation}
    \psi(t)=e^{-iH_{\mathrm{ho}}t/\hbar}\psi_0.
\end{equation}
The Schwartz assumption means that the wavefunction and its derivatives decrease rapidly at large distances.

Choose a smooth cutoff $\chi_L(Q)=\chi(Q/L)$, where $Q=(x,x_1,x_2)$, $0\leq\chi\leq1$, and $\chi\in C_c^\infty((-1,1)^3)$ equals one on $[-1/2,1/2]^3$. Thus $\chi_L$ leaves the wavefunction in the central region unchanged and makes it vanish before the walls. Define
\begin{equation}
    n_L=\|\chi_L\psi_0\|,
    \qquad
    \psi_L(t)=e^{-iH_Lt/\hbar}\frac{\chi_L\psi_0}{n_L}.
\end{equation}
Thus, $\psi_L(t)$ is obtained by evolving the restricted wavefunction. We extend $\psi_L$ by zero outside the box when comparing it with $\psi$. All wavefunction norms below are $L^2$ norms.

Set $\phi_L(t)=\chi_L\psi(t)$, which is equivalent to restricting the continuum wavefunction after its evolution. The continuum Schr\"odinger equation gives
\begin{equation}
    i\hbar\partial_t\phi_L =H_L\phi_L-[H_{\mathrm{ho}},\chi_L]\psi.
\end{equation}
Using this and the evolution of $\psi_L$, we can obtain
\begin{equation}
    \|\psi_L(t)-\phi_L(t)\| \leq |1-n_L| +\frac1\hbar\int_0^t \|[H_{\mathrm{ho}},\chi_L]\psi(s)\|\,ds.
\end{equation}
Since $\phi_L=\chi_L\psi$, the triangle inequality then gives
\begin{equation}
\begin{split}
    \|\psi_L(t)-\psi(t)\|\leq & |1-n_L|+\|(1-\chi_L)\psi(t)\|\\
    &\quad+\frac1\hbar\int_0^t \|[H_{\mathrm{ho}},\chi_L]\psi(s)\|\,ds.
\end{split}
\label{finite_box_error}
\end{equation}

We bound these three terms on a fixed interval $0\leq t\leq T$. For a Schwartz initial state, quadratic evolution ensures that $\psi(t)$ remains Schwartz for all times \cite[Corollary~5.3]{Wahlberg2022}. In particular, multiplication by a position coordinate or taking a first spatial derivative give square-integrable functions that depend continuously with time. Thus
\begin{equation}
\begin{gathered}
\||Q|\psi(t)\|^2 =\int_{\mathbb R^3}|Q|^2|\psi(Q,t)|^2\,dQ,\\
\|\nabla\psi(t)\|^2 =\frac{1}{\hbar^2} \left( \|p\psi(t)\|^2+ \|p_1\psi(t)\|^2+ \|p_2\psi(t)\|^2 \right)
\end{gathered}
\end{equation}
are finite and continuous functions of $t$. Consequently, on a closed finite time interval,
\begin{equation}
\begin{gathered}
    M_T:=\sup_{0\leq s\leq T}\||Q|\psi(s)\|<\infty,\\
    G_T:=\sup_{0\leq s\leq T}\|\nabla\psi(s)\|<\infty.
\end{gathered}
\end{equation}
These constants are related only to the continuum evolution of the fixed initial state, so they do not depend on the box size $L$.

The $1-\chi_L$ can be nonzero only where $|Q|\geq L/2$. Hence, it follows that
\begin{equation}
\begin{split}
    \|(1-\chi_L)\psi(t)\|^2 &\leq\int_{|Q|\geq L/2}|\psi(Q,t)|^2\,dQ\\
    &\leq\frac{4}{L^2} \int |Q|^2|\psi(Q,t)|^2\,dQ.
\end{split}
\end{equation}
Thus the removed tail is bounded by $2M_T/L$. The normalization error obeys the same bound:
\begin{equation}
    |1-n_L| \leq\|(1-\chi_L)\psi_0\| \leq\frac{2M_T}{L}.
\end{equation}

The remaining term gives
\begin{equation}
\begin{split}
    [H_{\mathrm{ho}},\chi_L]\psi =&-\frac{\hbar^2}{2m} \left[ 2\nabla\chi_L\cdot\nabla\psi +(\Delta\chi_L)\psi \right]\\
    &-i\hbar g \left[ x\,\partial_{x_1}\chi_L +x_2\,\partial_x\chi_L \right]\psi.
\end{split}
\end{equation}
Here $\nabla \equiv \begin{pmatrix}
    \p_x & \p_{x_1} & \p_{x_2}
\end{pmatrix}^T$ and $\Delta\equiv \p_x^2 + \p_{x_1}^2 + \p_{x_2}^2$. The chain rule gives
\begin{equation}
    \|\nabla\chi_L\|_\infty\leq\frac{C}{L},
    \qquad
    \|\Delta\chi_L\|_\infty\leq\frac{C}{L^2}.
\end{equation}
The derivatives vanish except in the transition region $\mc A_L$ between the central half-box and the walls. In this region, $|x|,|x_2|\leq L$, and the preceding tail estimate gives $\|\mathbf1_{\mc A_L}\psi(s)\|\leq2M_T/L$. Consequently,
\begin{equation}
    \|[H_{\mathrm{ho}},\chi_L]\psi(s)\| \leq C\left[ \frac{\hbar^2G_T}{mL} +\frac{\hbar^2}{mL^2} +\frac{\hbar|g|M_T}{L} \right].
\end{equation}
We used $\|\psi(s)\|=1$ in the second term. The constant $C$ depends only on the chosen cutoff.

Combining the three bounds gives
\begin{equation}
\begin{split}
    \varepsilon_L(T) &:=\sup_{0\leq t\leq T}\|\psi_L(t)-\psi(t)\|\\
    &\leq \frac{4M_T}{L} +CT\left[ \frac{\hbar G_T}{mL} +\frac{\hbar}{mL^2} +\frac{|g|M_T}{L} \right].
\end{split}
\end{equation}
Every term tends to zero as $L\to\infty$ at fixed $T$. The estimate holds for every finite $T$. However, because $M_T$ and $G_T$ can grow with $T$, the bound is not uniform as $T\to\infty$. Keeping the error small over a longer interval may require a larger box.

Now, from the wavefunction error, we can also bound the error in the reduced density matrix:
\begin{equation}
    \|\rho_{S,L}(t)-\rho_S(t)\|_1 \leq2\varepsilon_L(T).
\end{equation}
Here $\|\cdot\|_1$ is the trace norm.  Finally, the difference in the reduced Wigner function between the continuum and the finite-volume setting can be shown to be
\begin{equation}
    \|W_{S,L}(\cdot,t)-W_S(\cdot,t)\|_\infty \leq\frac{2}{\pi\hbar}\varepsilon_L(T).
\end{equation}
The reduced-state estimate follows from the pure-state trace-norm bound and contraction under partial trace. The Wigner estimate uses $\|W_\rho-W_\sigma\|_\infty \leq\|\rho-\sigma\|_1/(\pi\hbar)$. Both estimates hold for $0\leq t\leq T$.

We can therefore summarize the argument as follows: for a finite time and a prescribed accuracy, we can always choose the $L$ to be sufficiently large so that the continuum reduced Wigner function is reproduced throughout the interval. The Hamiltonian of this confined model is bounded below. The result concerns finite observation times.

\bibliography{references}

\end{document}